\documentclass[ACS,STIX1COL]{WileyNJD-v2}

\usepackage{color}

\usepackage{mathtools}
\usepackage{graphicx}
\usepackage{dcolumn}
\usepackage{array}
\usepackage{lipsum}
\usepackage{bm}
\usepackage{subfigure}
\usepackage{multirow}
\usepackage{tabularx}
\usepackage{amsmath}
\usepackage{braket}
\graphicspath{{plots/}}

\newcommand{\eps}[0]{\varepsilon}	

\DeclareRobustCommand{\orderof}{\ensuremath{\mathcal{O}}}
\renewcommand{\vec}[1]{\mathbf{#1}}

\newcommand{\beq}{\begin{equation}}
\newcommand{\eeq}{\end{equation}}
\newcommand{\bea}{\begin{eqnarray}}
\newcommand{\eea}{\end{eqnarray}}

\newcommand{\ofqw}{(q,\omega)}

\articletype{ORIGINAL ARTICLE}%

\received{xxx}
\revised{xxx}
\accepted{xxx}

\begin{document}

\title{\textit{Ab initio} results for the plasmon dispersion and damping of\\ the warm dense electron gas}

\author[1]{Paul Hamann}

\author[2]{Jan Vorberger}

\author[3]{Tobias Dornheim}

\author[4,5]{Zhandos A. Moldabekov}

\author[1]{Michael Bonitz*}

\authormark{Paul Hamann \textsc{et al}}

\address[1]{\orgdiv{Institut f\"ur Theoretische Physik und Astrophysik}, \orgname{Christian-Albrechts-Universit\"at zu Kiel}, \orgaddress{\state{Leibnizstra{\ss}e 15, 24098 Kiel}, \country{Germany}}}

\address[2]{\orgdiv{Helmholtz-Zentrum Dresden-Rossendorf} \orgname{}, \orgaddress{\state{Bautzner Landstra{\ss}e 400,  D-01328 Dresden}, \country{Germany}}}

\address[3]{\orgdiv{Center for Advanced Systems Understanding (CASUS)} \orgname{}, \orgaddress{\state{G\"orlitz}, \country{Germany}}}

\address[4]{\orgdiv{Institute for Experimental and Theoretical Physics}, \orgname{Al-Farabi Kazakh National University}, \orgaddress{\state{71 Al-Farabi str.,  
 050040 Almaty}, \country{Kazakhstan}}}
\address[5]{\orgname{Institute of Applied Sciences and IT}, \orgaddress{\state{40-48 Shashkin Str., 050038 Almaty}, \country{Kazakhstan}}}




\abstract{
Warm dense matter (WDM) is an exotic state on the border between condensed matter and dense plasmas. Important occurrences of WDM include dense astrophysical objects, matter in the core of our Earth, as well as matter produced in strong compression experiments. As of late x-ray Thomson scattering has become an advanced tool to diagnose WDM. The interpretation of the data requires model input for the dynamic structure factor $S(q,\omega)$ and the plasmon dispersion $\omega(q)$. Recently the first \textit{ab initio} results for $S(q,\omega)$ of the homogeneous warm dense electron gas were obtained from path integral Monte Carlo  simulations,  [Dornheim \textit{et al.}, Phys. Rev. Lett. \textbf{121}, 255001 (2018)]. Here, we analyse the effects of correlations and finite temperature on the dynamic dielectric function and the plasmon dispersion. Our results for the plasmon dispersion and damping differ significantly from the random phase approximation and from earlier models of the correlated electron gas. Moreover, we show when commonly used weak damping approximations break down and how the method of complex zeros of the dielectric function can solve this problem for WDM conditions.
}



\maketitle



\section{Introduction}\label{sec1}

There is growing interest in warm dense matter (WDM) \cite{graziani-book, dornheim_physrep_18}---an extreme state that occurs, e.g., in astrophysical objects~\cite{militzer_massive_2008,saumon_the_role_1992,schlanges-etal.95cpp,bezkrovny_pre_4}, in the core of our Earth \cite{hausoel_natcom_17}, in laser compression experiments \cite{Ernstorfer1033}, and on the pathway towards inertial confinement fusion~\cite{hu_militzer_PhysRevLett.104.235003,hurricane_inertially_2016}.
WDM is a complicated state due to an intricate interplay of many effects including quantum degeneracy and exchange of the electrons at finite temperatures, electronic and ionic correlations, including wide angle scattering, phase transitions, and partial ionization. The simultaneous occurrence of these effects makes the experimental and theoretic analysis of WDM extremely challenging. Among the experimental diagnostics, x-ray Thomson scattering (XRTS) has been established as an accurate and highly promising tool \cite{redmer_glenzer_2009}. XRTS measures the dynamic structure factor of all the electrons in the system
\begin{align}
S(q,\omega)=  \frac{1}{\pi n \tilde v(q)}\frac{1}{e^{-\beta\hbar\omega}-1} \mbox{Im}\frac{-1}{\epsilon(q,\omega)}\,,     
\label{eq:s-def}
\end{align}
and yields information, among others, on the plasmon spectrum, density, temperature, and chemical composition. In Eq.~(\ref{eq:s-def}), $\beta=1/k_BT$, $n$ is the density,  $\tilde v(q)=4\pi e^2/q^2$ is the Fourier transform of the Coulomb potential, and $\epsilon(q,\omega)$ denotes the dynamic dielectric function. In order to use XRTS as diagnostics, input from models for the dynamic structure factor, i.e., dynamic dielectric function $\epsilon\ofqw$ is required. Thus, the accuracy of the diagnostics of WDM crucially depends on the quality of the available models.

This concerns, in particular, the dispersion $\omega(q)$ and damping $\gamma(q)$ of collective excitations. 
There has been extensive recent theoretical work on this subject which includes chemical models, sum rule models~\cite{arkhipov_prl17, arkhipov_cpp18}, local field correction (LFC) theory \cite{fortmann_pre_10,Dornheim_ESA}, time-dependent density functional theory (TDDFT) \cite{baczewski_prl_16}, quantum kinetic theory \cite{bonitz_qkt, bonitz_cpp18}, non-equilibrium Green's functions \cite{green-book, kwong_prl_00,schluenzen_cpp_18}, and quantum hydrodynamics \cite{moldabekov_pop15,zhandos_pop18,bonitz_pop_19,zhandos_cpp_19}. 
%
Standard assumptions that are used include the
Chihara decomposition in multi-component plasmas (this incorporates the Born-Oppenheimer approximation and leads to the description of the response of the free electrons by an electron gas model) \cite{Chihara_1987,wunsch_epl2011}, the decoupling of longitudinal and transverse modes, and the weak damping approximation, e.g. Ref.~\cite{alexandrov-book_84}. Due to the complexity of WDM, the influence of each of these approximations is often difficult to quantify and reliable predictive models are still missing.

It is, therefore, useful to disentangle these effects by using simpler but well-defined model systems. One such system is the uniform electron gas (UEG) at finite temperature (jellium). This is an example of an one-component system that constitutes an important test case for theory, as it allows one to focus on the treatment of quantum, correlation and finite temperature effects of the electrons and to benchmark models against first principle simulations. The plasmon dispersion of the UEG has been studied for many decades starting with the works of Bohm, Gross, Pines and Ferrell for metals \cite{bohm-gross_49, bohm-pines-4,ferrell_57} who developed a quantum mean field theory within the random phase approximation (RPA); for useful parametrizations see Ref~\cite{arista-brandt_84}.  Correlation effects were taken into account via local field corrections, e.g. in Refs.~\cite{stls_68,vashishta-singwi_72,Tanaka_1985,dandrea_86}. Recent applications to WDM include the analysis of experiments on beryllium and boron \cite{neumayer_prl_10, thiele_pre_08,fortmann_pre_10}. 

However, the accuracy of these model results for the plasmon dispersion and damping, in particular under WDM conditions, is not known. Therefore, it is highly desirable to develop simulations that avoid approximations regarding correlation, quantum and finite temperature effects. The most accurate approach available is path integral Monte Carlo (PIMC) which, however, also faces fundamental difficulties such as the fermion sign problem~\cite{loh_prb_90,troyer_prl_05,dornheim_pre_19} and the complicated access to frequency dependent observables. 
These problems were overcome by the combination of configuration PIMC (CPIMC) \cite{schoof_cpp11,schoof_prl15}, that is exact at strong degeneracy but exhibits a sign problem at weak degeneracy (low density), and of permutation blocking PIMC (PB-PIMC) \cite{dornheim_njp15,dornheim_jcp15,dornheim_cpp_19}. Thus \textit{ab initio} simulations of jellium now cover the entire range of WDM parameters \cite{dornheim_prb16,groth_prb16} given by $0.1 \lesssim\Theta, r_s \lesssim 10$, where $\Theta=k_BT/E_F$ and $r_s=\bar r/a_B$, with $E_F$ and $a_B$ denoting the Fermi energy and the Bohr radius. 
Finally, an accurate extrapolation to the thermodynamic limit was achieved in Ref.~\cite{dornheim_prl16} and the connection to the ground state was realized in Ref.~\cite{groth_prl17}, where also an accurate parametrization for the exchange-correlation free energy was reported. For an overview on the results and comparisons with earlier models and simulations, see Ref.~\cite{dornheim_pop17,dornheim_physrep_18}.

PIMC simulations can also be used to obtain dynamic quantities such as the dynamic structure factor $S(q,\omega)$
from an analytical continuation of the intermediate scattering function (density correlation function) $F(q,t)$ evaluated at imaginary times $\tau\in[0,\beta]$.
The scattering function is related to the dynamic structure factor $S(q,\omega)$ by a Laplace transform
\begin{align}
\label{eq:S_F}
F(q,\tau) = \int_{-\infty}^\infty \textnormal{d}\omega\ S(q,\omega)\, e^{-\tau\omega} \, .
\end{align}
This is known to be an ill-posed problem that has occasionally been tackled using  maximum entropy methods, see e.g. Ref.~\cite{boninsegni_prb18} and references therein.
Recently it was found that a stochastic sampling of the LFCs $G(q,\omega)$ allows to very well reconstruct the imaginary time density response function and thus the dynamic structure factor, because additional exact constrains on the LFC makes the procedure very efficient and accurate~\cite{dornheim_prl_18}. Extensive further studies were reported in Ref. \cite{groth_prb_19}. It was also noted in Ref.~\cite{dornheim_prl_18}, that in many cases the static LFC, $G(q)=G(q,\omega=0)$ is sufficiently accurate to recover the dynamic structure factor. For this purpose, a neural net representation for $G(q)$ was constructed in Ref.~\cite{dornheim_jcp_19} that is based on \textit{ab initio} simulation data. Access to $G(q)$ or even $G(q,\omega)$ allows for systematic extensions of the QMC based \textit{ab initio} approach to other dynamic quantities such as the density response function $\chi(q,\omega)$ \cite{bonitz_pop_19}, the dielectric function, and the dynamic conductivity \cite{hamann_prb_20}.

For the plasmon dispersion, the key quantity is the dynamic dielectric function $\epsilon\ofqw$ 
\begin{equation}
    \epsilon(q,\omega) = 1 - \frac{\tilde v(q)\Pi^{\rm RPA}(q,\omega)}{1+\tilde v(q) G(q,\omega) \Pi^{\rm RPA}(q,\omega)}\,,     
    \label{eq:eps-G}
\end{equation}
where $\Pi^{\rm RPA}$ is the polarization function in random phase approximation (RPA, Lindhard). Setting $G\to 0$, recovers the RPA dielectric function. On the other hand, Eq.~(\ref{eq:eps-G}) indicates that, with \textit{ab initio} input for $G(q,\omega)$ also \textit{ab initio} results for the dielectric function of correlated electrons under WDM conditions are becoming available.

The goal of this paper is to study the dynamic dielectric function in more detail with a focus on its zeroes because they determine the plasmon dispersion, $\omega(q)$, and damping, $\gamma(q)$, of correlated electrons. In particular, 
\begin{enumerate}
    \item We present a detailed analysis of the wave number dispersion, $\omega(q)$, at finite temperature. Starting with RPA, we review various analytical models and find that they exhibit significant deviations from the numerical result. We also present a novel analytical parametrization at finite temperature for $\omega(q)$ in RPA. 
    \item We present results for the plasmon dispersion and damping that follow from the PIMC results for the local field correction $G(q,\omega)$ and compare the results to previous studies. 
    \item We carefully test the validity of the commonly used dispersion relation, Re $\epsilon(q,\omega)=0$. Since in XRTS experiments under WDM conditions the plasmon damping is not necessarily small, this relation has to be questioned. Therefore, we perform an analytical continuation of the retarded dielectric function to complex frequencies \cite{bonitz_qkt}. We present results for the RPA dielectric function and for Eq.~(\ref{eq:eps-G}) and observe significant deviations from the common approach based on the real part of $\epsilon$. This has important implications for the correct interpretation of the XRTS measurements of warm dense matter.
\end{enumerate}

The paper is organized as follows: In Sec.~\ref{s:pimc}, we summarize the main ideas of our PIMC approach to the dielectric function that is based on the reconstruction of the dynamic local field correction. There, we also present a discussion of the longitudinal plasmon dispersion and of the analytical continuation. In Sec.~\ref{s:results}, we present our \textit{ab initio} simulation results for the local field correction, the dielectric function, and the plasmon dispersion. We conclude with a summary and outlook in Sec.~\ref{s:summary}.


\section{Path integral Monte Carlo approach to the plasmon dispersion}\label{s:pimc}
\subsection{Path integral Monte Carlo}\label{ss:pimc}

The basic idea of the standard path integral Monte Carlo method~\cite{cep} is to stochastically evaluate the thermal density matrix 
\begin{eqnarray}\label{eq:rho}
\rho(\mathbf{R}_a,\mathbf{R}_b,\beta) = \bra{\mathbf{R}_a} e^{-\beta\hat H} \ket{\mathbf{R}_b}\ ,
\end{eqnarray}
in coordinate space, with $\mathbf{R}=(\mathbf{r}_1,\dots,\mathbf{r}_N)^T$ containing the coordinates of all $N$ particles and $\beta=1/k_\textnormal{B}T$ being the ususal inverse temperature.
As a direct evaluation of $\rho(\mathbf{R}_a,\mathbf{R}_b,\beta)$ is not possible, one performs a Trotter decomposition~\cite{deRaedt_Trotter}, and the final result for the partition function $Z$ is given as the sum over all closed paths of particle coordinates in the imaginary time $\tau\in[0,\beta]$, see Refs.~\cite{dornheim_pre_19,dornheim_jcp_19} for details.

We note that this formulation in the imaginary time is particularly convenient in the context of the present work, as it allows for a straightforward computation of imaginary-time correlation functions, such as the density autocorrelation function
\begin{eqnarray}\label{eq:define_F}
F(q,\tau) = \frac{1}{N}  \braket{\hat\rho(q,\tau)\hat\rho(-q,0)}\ .
\end{eqnarray}

All PIMC data presented in this work have been obtained without any nodal restrictions~\cite{Ceperley1991} in Eq.~(\ref{eq:rho}). Therefore, the simulations are computationally demanding due to the fermion sign problem (see Ref.~\cite{dornheim_pre_19} for a review article), but exact within the given error bars.
Moreover, we use a canonical adaption~\cite{mezza} of the worm algorithm introduced by Boninsegni \textit{et al.}~\cite{boninsegni1,boninsegni2}.


\subsection{Stochastic sampling of the dynamic LFC}\label{ss:sampling}

The numerical inversion of Eq.~(\ref{eq:S_F}) is a notoriously hard problem~\cite{Jarrell_Review_1996}. Solutions for $S(q,\omega)$ are, in general, not unique as the information contained in the PIMC data for $F(q,\tau)$ does not fully determine the DSF.
To overcome this obstacle, Dornheim and co-workers~\cite{dornheim_prl_18,groth_prb_19,dornheim_HEDP_2020} have introduced a stochastic sampling scheme for the dynamic local field correction, which automatically satisfies a number of exact constraints on $G(q,\omega)$ and, in this way, sufficiently constraints the space of possible solutions for $S(q,\omega)$.

The basic workflow of this method is as follows~\cite{dornheim_prl_18,groth_prb_19}: 1) Generate a random trial solution for Im$G(q,\omega)$ that already incorporates a number of well-known exact relations. 2) Use the Kramers-Kronig relations~\cite{giuliani2005quantum} to compute the corresponding real part, Re$G(q,\omega)$. 3) Use both parts to compute a corresponding trial solution for $\chi(q,\omega)$ [or, equivalently, $\epsilon(q,\omega)$, cf.~Eq.~(\ref{eq:eps-G})]. 4) Use the fluctuation--dissipation theorem (\ref{eq:s-def}) to compute the trial solution for $S(q,\omega)$. 5) Insert $S(q,\omega)$ into Eq.~(\ref{eq:S_F}) and measure the deviation to the PIMC data for $F(q,\tau)$ for all $\tau$-points. Only those $G(q,\omega)$ which lead to an imaginary-time density--density correlation function in agreement to the PIMC data constitute valid solutions.



\subsection{LFCs and the dynamic dielectric function}\label{ss:epsilon}

Having obtained \textit{ab initio} results for the dynamic local field correction, it is straightforward to obtain the dynamic dielectric function via Eq.~(\ref{eq:eps-G}). 
This can be rewritten in terms of the polarization function $\Pi$
\begin{align}\label{eq:eps-def}
    \epsilon(q,\omega) &= 1 - \tilde v(q)\Pi(q,\omega),
\end{align}
where
\begin{align}
    \Pi(q,\omega)&= \frac{\Pi^{\rm RPA}(q,\omega)}{1+\tilde v(q)G(q,\omega)\Pi^{\rm RPA}(q,\omega)}\,.
    \label{eq:pi-def}
\end{align}
In the mean field limit, $G \to 0$, and we recover the RPA dielectric function
\begin{align}
    \epsilon^{\rm RPA}(q,\omega) &= 1 - \tilde v(q)\Pi^{\rm RPA}(q,\omega)\,.
    \label{eq:eps_rpa}
\end{align}
As already mentioned, an important approximation is obtained by replacing $G(q,\omega)$ with its static limit, $G(q,\omega) \rightarrow G(q,0) = G(q)$ in Eq.~(\ref{eq:eps-G}) or Eq.~(\ref{eq:pi-def}). This is still a dynamic dielectric function which will be denoted $\epsilon^{\rm SLFC}(q,\omega)$, while the full dynamic result will be called $\epsilon^{\rm DLFC}(q,\omega)$.
Comparing results for the dynamic structure factor revealed, that the static approximation is accurate for $r_s\lesssim 4$, for all wave numbers~\cite{dornheim_prl_18}.


\subsection{Longitudinal plasmon dispersion}\label{ss:dispersion}

The existence of longitudinal collective plasma oscillation follows from Maxwell's equations which, after Fourier transform, reduce to a wave equation for the Fourier components of the electric field strength~\cite{alexandrov-book_84},
\begin{align}
    \left\{q^2\delta_{\alpha,\beta}-q_\alpha q_\beta -\frac{\omega^2}{c^2}\epsilon_{\alpha\beta}(\textbf{q},\omega) \right\}E_\beta(\textbf{q},\omega)=0\,,
    \label{eq:maxwell-eq}
\end{align}
which are three coupled equations for the complex cartesian components $E_\alpha(\textbf{q},\omega), \,\; \alpha, \beta=x,y,z$ with $\epsilon_{\alpha\beta}$ being the dielectric tensor of the medium. Non-trivial solutions for the field strength exist if the determinant of the matrix in the braces vanishes.
In an isotropic medium, such as a plasma, the dielectric tensor has only two non-vanishing components -- the longitudinal, [$\epsilon(q,\omega)$], and the transverse, [$\epsilon^{tr}(q,\omega)$] dielectric functions -- which have to vanish simultaneously. Here we are only interested in the longitudinal part as it yields the longitudinal plasmon dispersion and damping. Vanishing of the determinant of the matrix in Eq.~(\ref{eq:maxwell-eq}) then reduces to the dispersion relation
\begin{align}
    \epsilon[\hat \omega(q),q] = 0\,,
    \label{eq:dispersion}
\end{align}
where $\hat \omega(q)$ is the plasmon frequency for wavenumber $q$ which is, in general, a complex function. This means the solution of Maxwell's equation in the plasma, following a (possibly random) excitation with wave number q have a time-dependence that is governed by the solution of Eq.~(\ref{eq:dispersion}), 
$E(q,t) \sim e^{-i\hat\omega(q)t}$. In thermodynamic equilibrium, this solution has to vanish in the long time limit, which requires Im$\,\hat \omega(q)<0$.

In case of weak damping, this solution can be approximated by the roots of the real part, and the plasmon damping $\gamma(q)$ follows in perturbation theory~\cite{bonitz_qkt},
\begin{align}
    0 &= \mbox{Re}\, \epsilon[\omega(q),q] ,
    \label{eq:real-dispersion}\\
    \gamma(q) &= \frac{\mbox{Im}\, \epsilon[\omega(q),q]}{\frac{\partial}{\partial \omega}\mbox{Re}\, \epsilon[\omega(q),q]}\,,\quad |\gamma(q)|\ll\omega(q)\,. 
    \label{eq:real-damping}
\end{align}
%
\subsubsection{Weak damping approximation}\label{ss:results-dispersion-wd}

Let us now turn to the dispersion of collective plasmon oscillations which, for weak damping, is derived from Re $\epsilon^{\rm RPA}=0$. In a plasma in 3D in the limit $q\to 0$, the dispersion starts at $\omega=\omega_p$. For finite $q$, there appear corrections that involve even powers of $q$. There is a large body of work for classical and quantum plasmas at zero and finite temperature in the absence of correlation effects. Let us summarize some common approximations that include the terms of order $q^2$ and $q^4$ (a derivation of the full RPA results at finite temperature is presented in the appendix):
\begin{enumerate}
    \item The first result for the $q^2$ term of a classical plasma is due to Bohm and Gross \cite{bohm-gross_49},
    \begin{equation}
        \frac{\omega^2(q)}{\omega_p^2} = 1  + \frac{v^2_{\rm th}}{\omega^2_p}\, q^2\,,
        \label{eq:bohm-gross}
    \end{equation}
    where $v^2_{\rm th}=3 k_BT/m$ is the thermal velocity.
    \item In a degenerate quantum plasma at $T=0$, this dispersion is replaced by  \cite{alexandrov-book_84}
        \begin{equation}
        \frac{\omega^2(q)}{\omega_p^2} = 1  + \frac{3}{5}\frac{v^2_{\rm F}}{\omega_p^2}\, q^2\,.
        \label{eq:fermi-dispersion}
    \end{equation}
    \item The first account of the $q^4$ corrections to the dispersion for an ideal Fermi gas at $T=0$ is due to Bohm and Pines \cite{bohm-pines-3}, and the result was subsequently improved  by Ferrell \cite{ferrell_57} who reported 
        \begin{align}
        \frac{\omega^2(q)}{\omega_p^2} &= 1  + \frac{\langle v^2\rangle_0}{\omega_p^2}\, q^2 + \left\{ \frac{\left(\Delta v^2_0\right)^2}{\omega^2_p} + \frac{\hbar^2}{4m^2}\right\}\frac{q^4}{\omega_p^2}
        \label{eq:ferrell-dispersion}\\\nonumber
        & = 1+ \frac{3}{5}4\gamma^2\frac{q^2}{q^2_F} + \gamma^2\left\{16\gamma^2\left(\frac{12}{175}\right)^2 + 1
        \right\}
        \left(\frac{q}{q_F}\right)^4\,,
    \end{align}
    where $\gamma=E_F/\hbar \omega_p$, $\langle v^2\rangle_0 = \frac{3}{5}v_F^2$ and $\Delta v^2_0 = [\langle v^4 \rangle_0 - \langle v^2 \rangle_0^2]^{1/2}$, and the subscript ``0'' indicates the average with the ground state Fermi function. Note that the term with $(\Delta v^2_0)^2$ [i.e. with $16\gamma^2$] is not present in Ref.~\cite{bohm-pines-3} and most other works.
    \item An extension of the ideal Fermi gas parametrization to finite temperatures was reported by Arista and Brandt \cite{arista-brandt_84} and, more recently, by Thiele \textit{et al.} ~\cite{thiele_pre_08}
        \begin{align}
        \frac{\omega^2(q)}{\omega_p^2} &= 1  + \frac{\langle v^2\rangle}{\omega^2(q)}\, q^2 + \left\{ \frac{\langle v^4\rangle}{\omega^2(q)} + \frac{\hbar^2}{4m^2}\right\}\frac{q^4}{\omega^2(q)}\,,
        \label{eq:thiele-dispersion}
    \end{align}
who then replaced, in the denominator, $\omega(q)\to \omega_p$.    
For finite temperatures on the order of $\theta=1$, the authors of Ref.~\cite{thiele_pre_08} neglected the $\langle v^4\rangle$ term and  proposed the following parametrization:
        \begin{align}
        \frac{\omega^2(q)}{\omega_p^2} &= 1  + \frac{ v^2_{\rm th}}{\omega^2_p}(1+0.088 \,\chi)\, q^2 +  
        \frac{\hbar^2}{4m^2}\frac{q^4}{\omega^2_p}\,,
        \label{eq:thiele-dispersion-theta1}
    \end{align}
    with the degeneracy parameter $\chi=n\Lambda^3=nh^3(2\pi mk_BT)^{-3/2}$. 
\item The finite temperature RPA dispersion (\ref{eq:thiele-dispersion}) can be further improved if the terms $\omega(q)$ in the denominator are not replaced by $\omega_p$ but, instead, the full result is used iteratively:
 \begin{align}
        \frac{\omega^2(q)}{\omega_p^2} &= 1  + \frac{\langle v^2\rangle}{\omega_p^2}\, q^2 + \left\{ \frac{\left(\Delta v^2\right)^2}{\omega^2_p} + \frac{\hbar^2}{4m^2}\right\}\frac{q^4}{\omega_p^2}\,,
        \label{eq:RPA_dispersion-finite-t}
    \end{align}
where the velocity moments are computed with the finite temperature Fermi function. This gives the most accurate result for the $q^4$ coefficient (see Appendix). Evaluating the Fermi integrals we find the following  parametrization of the dispersion, where, for the coefficients, very accurate analytical expressions are presented in the Appendix,
 \begin{align}
        \frac{\omega^2(q)}{\omega_p^2} &= 1  + B_2(r_s,\theta)\left(\frac{q}{q_F}\right)^2  + B_4(r_s,\theta)\left(\frac{q}{q_F}\right)^4\,,
        \label{eq:RPA_dispersion-finite-t-param}
    \end{align}


\end{enumerate}

\begin{figure}
    \centering
    \includegraphics[width=0.8\textwidth]{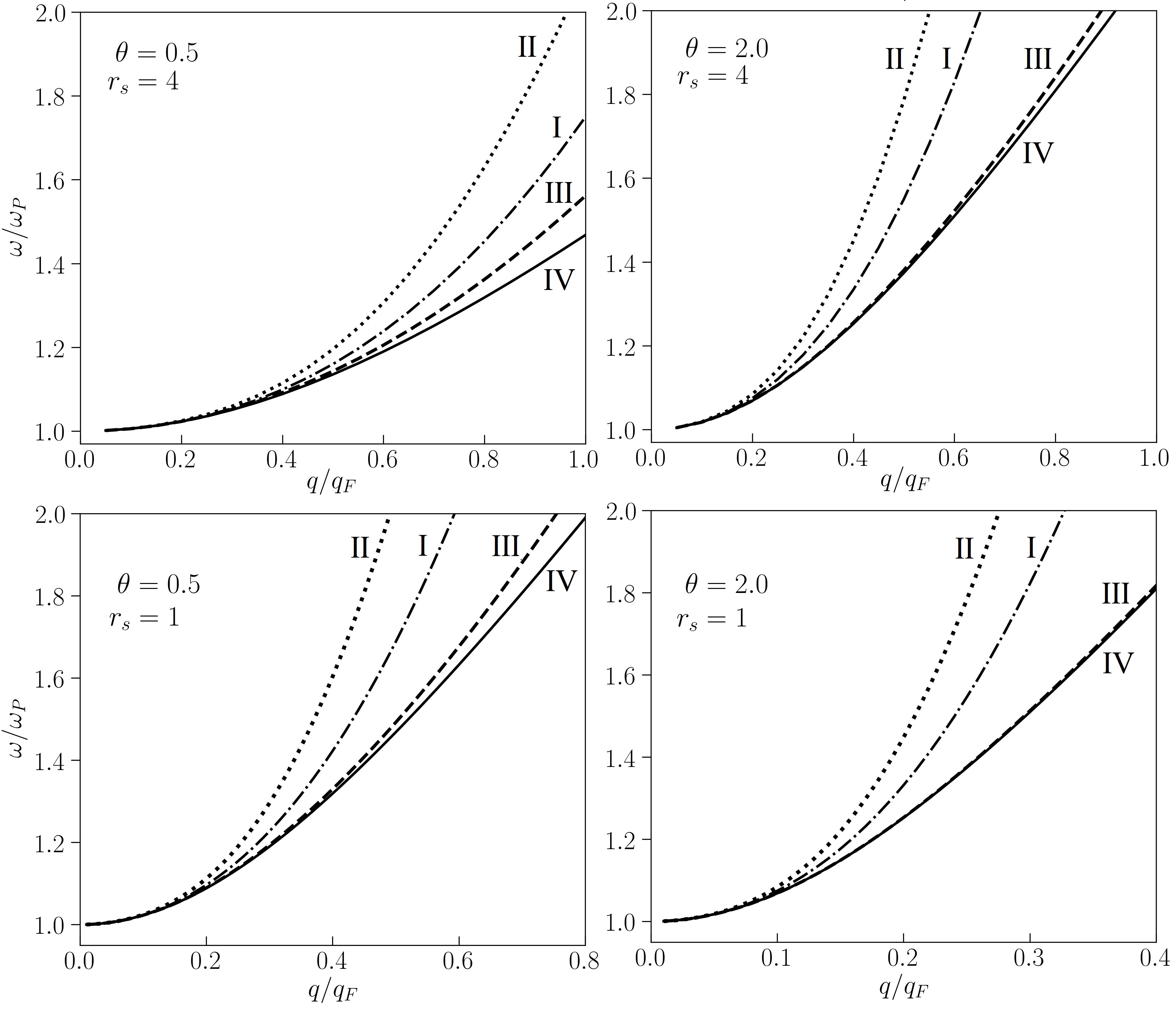}
    \caption{Comparison of various approximations for the RPA plasmon dispersion derived from Re $\epsilon^{\rm RPA}=0$, at $r_s=4$ (top), and $r_s=1$ (bottom), for two temperatures, $\theta=0.5$ (left column) and $\theta=2$ (right). Dash-dotted line (I): full result,  
    Eq.~(\ref{eq:RPA_dispersion-finite-t}); dotted line (II): result of 
    Eq.~(\ref{eq:thiele-dispersion});
    dashed line (III): neglecting the term proportional to $\left(\Delta v^2\right)^2$ in Eq.~(\ref{eq:RPA_dispersion2}), which is equivalent to Eq.~(\ref{eq:thiele-dispersion-theta1});     Solid line (IV): neglecting  $\orderof\left( \tilde{q}^{\,4}\right)$ terms in Eq.~(\ref{eq:RPA_dispersion-finite-t}). 
    }
    \label{fig:RPA_dispersion}
\end{figure}
The result (\ref{eq:RPA_dispersion-finite-t-param}) is shown in Fig.~\ref{fig:RPA_dispersion} by line ``I'' and is the most accurate RPA result in the weak damping approximation for small $q$. The result that follows when the term $(\Delta v^2)^2$ is replaced by $v^4$ is shown by line ``II''. Eq.~(\ref{eq:thiele-dispersion-theta1}) leads to the line labelled "III". It follows from neglecting $(\Delta v^2)^2$. The approximation that neglects, in Eq.~(\ref{eq:RPA_dispersion-finite-t}), all terms of order $q^4$, is shown by the line ``IV''. Note that all of these approximations do not take into account that the dispersion relation Re $\epsilon=0 $ has solutions only for a finite range of wavenumbers.

\subsubsection{Beyond weak damping}
If, however, the damping is not small, it is straightforward to improve approximations (\ref{eq:real-dispersion}) and (\ref{eq:real-damping}) by extending the Taylor expansion of the complex dispersion relation (\ref{eq:dispersion}) to terms that are second order in $|\gamma|/\omega$. The next order result has the form of two coupled equations (for details, see Appendix B)
\begin{align}
    0&= \text{Re}\,\epsilon(\omega) - \frac{\gamma^2}{2!}\text{Re}\,\epsilon''(\omega)
     +\gamma \text{Im}\,\epsilon'(\omega) \,,
\nonumber     
 \\
    0&= \gamma \text{Re}\,\epsilon'(\omega) 
 - \text{Im}\,\epsilon(\omega) \,.
 \label{eq:disp-expansion-re-im-parts}
\end{align}
This approximation improves the accuracy of the dispersion in the case of moderate damping, as we will demonstrate in Sec.~\ref{ss:results-dispersion}.

If the damping is large, the above Taylor expansion will fail, and we have to return to the full condition (\ref{eq:dispersion}). We rewrite it explicitly in terms of real and imaginary parts
\begin{align}
    \mbox{Re}\,\epsilon[\hat\omega(q),q] &=     \mbox{Im}\,\epsilon[\hat\omega(q),q] = 0,
    \label{eq:complex-dispersion}
    \\
    \hat\omega(q) &= \omega(q) - i \gamma(q)\,,
    \label{eq:c-frequency}
\end{align}
where we introduced the complex frequency $\hat \omega$. [Note that we defined $\gamma = -{\rm Im}\,\hat\omega$, which is positive in an isotropic equilibrium plasma, as shown above  and, moreover also in an isotropic plasma out of equilibrium \cite{bonitz94pp}.] The pair of equations (\ref{eq:complex-dispersion}), if solved for several points in the complex frequency plane, yields the dispersion and damping simultaneously, cf. Sec.~\ref{ss:results-dispersion}. The relations (\ref{eq:complex-dispersion}) imply that the retarded dielectric function has been analytically continued into the complex frequency plane which 
 has occasionally been done for the  mean field approximation \cite{bonitz-etal.93prl,bonitz-etal.94pre,vladimirov_ufn_2011}. Here we will report results for the RPA dielectric function, Eq.~(\ref{eq:eps_rpa}) and for the correlated dielectric function, using the static LFC.

\subsubsection{Analytic continuation of the dynamic dielectric function}\label{ss:ac}
Taking Eq.~(\ref{eq:dispersion}) or Eqs.~(\ref{eq:complex-dispersion}) seriously requires the knowledge of the retarded dielectric function in the complex plane. Therefore, it needs to be analytically continued from the real axis into the complex plane. The analytic continuation of the whole dielectric function is based on the complex continuation of the finite-temperature Lindhard polarization function.
For real frequencies, the retarded and advanced polarization functions are given by
\begin{equation}
  \Pi^{R/A}(\vec{q},\omega) = \int\frac{d\vec{p}}{(2\pi)^3} \frac{f\left(E_{\vec{p}}\right) - f\left(E_{\vec{p}+\vec{q}}\right)}{E_{\vec{p}} - E_{\vec{p}+\vec{q}}+\omega \pm i\delta}, \qquad \delta\rightarrow 0^+\,.
\nonumber
\end{equation}
Starting from the spectral function
\begin{align}
\hat{\Pi}(\vec{q},\omega) &= \frac{1}{i} \left\{ \Pi^R(\vec{q},\omega+i\delta) - \Pi^A(\vec{q},\omega-i\delta) \right\} 
\nonumber\\
  &= \int\frac{d\vec{p}}{(2\pi)^3} \left\{ f\left(E_{\vec{p}}\right) - f\left(E_{\vec{p}+\vec{q}}\right) \right\} \, \delta\left[\omega +  E_{\vec{p}} - E_{\vec{p}+\vec{q}}\right]\,,
  \label{eq:spectral}
\end{align}
we obtain the analytic continuation to the complex frequency plain using Cauchy's integral formula
\begin{align}
\Pi^{R/A} (\vec{q},z) &= \int\frac{d\omega}{2\pi} \frac{\hat{\Pi}(\vec{q},\omega)}{\omega-z} 
= \int\frac{d\vec{p}}{(2\pi)^3} \frac{f(E_{\vec{p}}) - f(E_{\vec{p}+\vec{q}})}{E_{\vec{p}} - E_{\vec{p}+\vec{q}}+z}\,,
\label{eq:response}
\end{align}
yielding $\Pi^R$ in the upper and $\Pi^A$ in the lower half-plain.

As discussed above, cf. Eq.~(\ref{eq:complex-dispersion}), collective modes follow from zeroes of the retarded dielectric function, which may appear in the lower half-plane, where the retarded polarization function is given by
\begin{equation}
\tilde{\Pi}^R(\vec{q},z) = \Pi^A(\vec{q},z) - 2\pi i\, \hat{\Pi}(\vec{q},z)\,.
\nonumber
\end{equation}
In order to evaluate the spectral function $\hat{\Pi}$ at complex frequencies, the integration in Eq.~(\ref{eq:spectral}) has to be performed first, yielding:
\begin{align}
i\hat{\Pi}(q,\omega) &=
 \frac{i}{q}
  \int\limits_{p_-}^\infty
    dp\, p 
    \left\{ f(p^2/2) - f(p^2/2 + \omega)  \right\} 
    = \frac{i}{k\beta} \ln
\frac{1+\exp\left[ \beta\mu - \beta \left(\frac{k}{2} + \frac{\omega}{2k}\right)^2\right]}{1+\exp\left[ \beta\mu - \beta \left(\frac{k}{2} - \frac{\omega}{2k}\right)^2\right]} \,.
\end{align}
Using this result, the replacement $\omega \to z=\hat\omega$ can be made, and the analytic continuation of $\Pi^R$ to the lower half-plane can be carried out. The analytic continuation of the real and imaginary part of the RPA dielectric function follows from this easily. These results will be used in our numerical analysis of the plasmon dispersion in Sec.~\ref{ss:results-dispersion}. 

To extend this result beyond the weak coupling case, we have to perform an analytic continuation of the correlated dielectric function. This problem is solved in the following way. Our starting point is the -- formally exact -- relation (\ref{eq:eps-G}), which includes correlation effects via our \textit{ab initio} data for the dynamic local field correction $G(q,\omega)$. Using the static approximation $G(q,0)$, this expression is easily continued (we denote the analytic continuation by a ``hat'')
\begin{align}
    \hat \epsilon^{\rm \,SLFC}(q,\hat \omega) = 1 - \frac{\tilde v(q)\tilde \Pi^{\rm RPA}(q,\tilde \omega)}{1+\tilde v(q) G(q) \tilde \Pi^{\rm RPA}(q,\hat \omega)}, \label{eq:complex-eps-Gstatic}
\end{align}
because it only involves the analytic continuation of the Lindhard function.


\section{Numerical Results}\label{s:results}


\subsection{\textit{Ab initio} dynamic dielectric function}\label{ss:results-epsilon}
%
\begin{figure}
    \centering
    \includegraphics[width=0.45\linewidth]{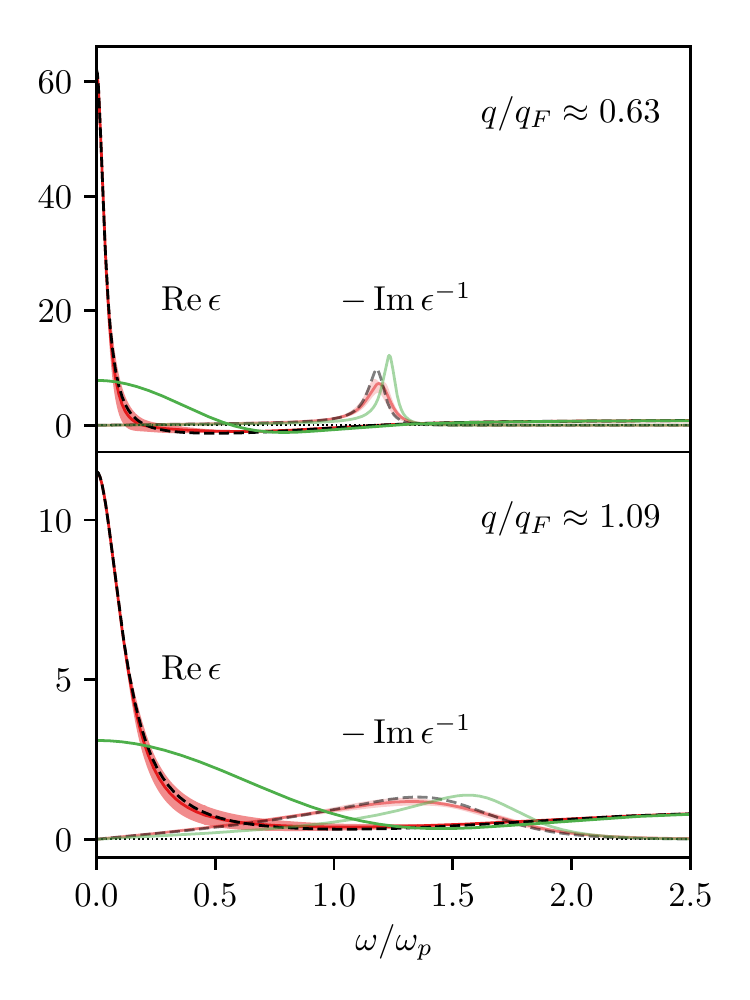}
    \includegraphics[width=0.45\linewidth]{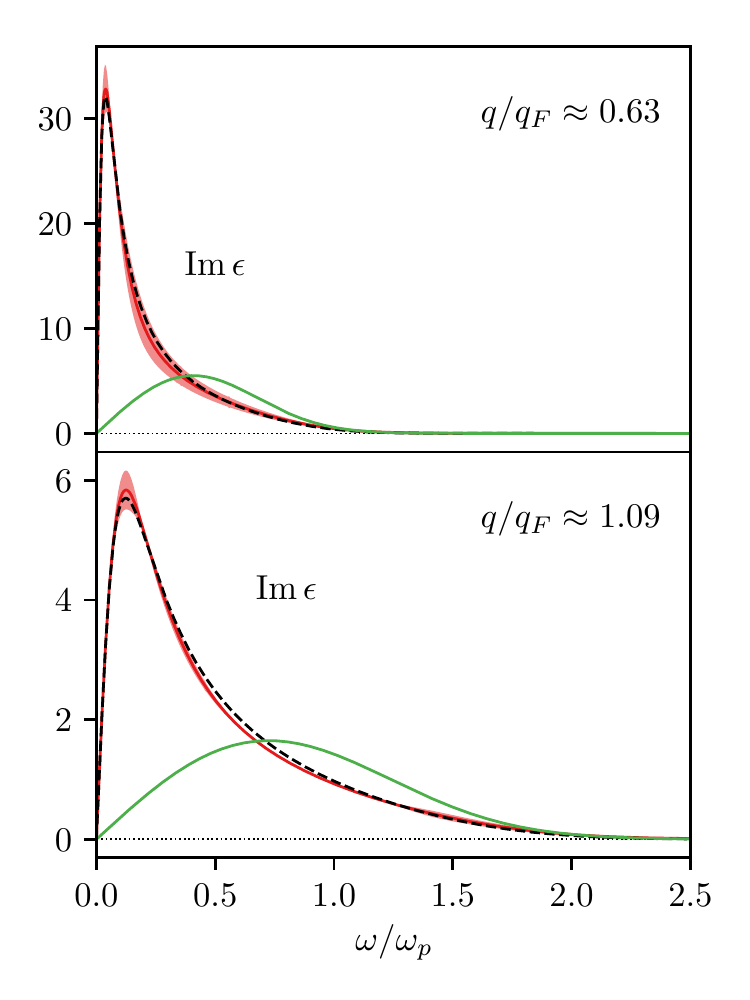}
    \includegraphics[width=0.45\linewidth]{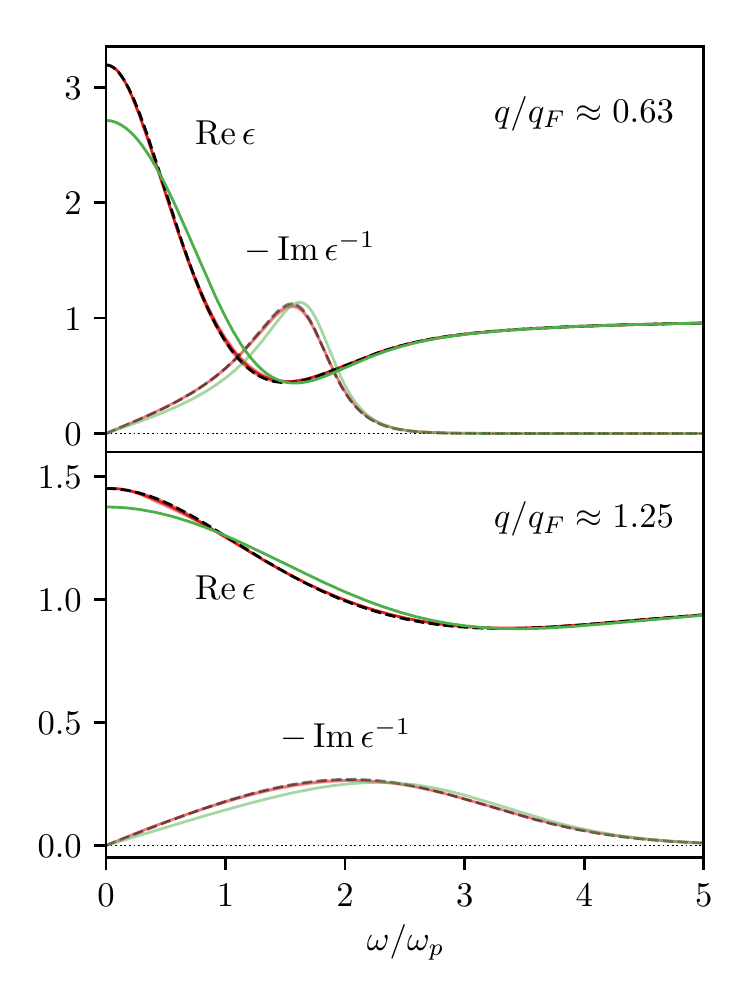}
    \includegraphics[width=0.45\linewidth]{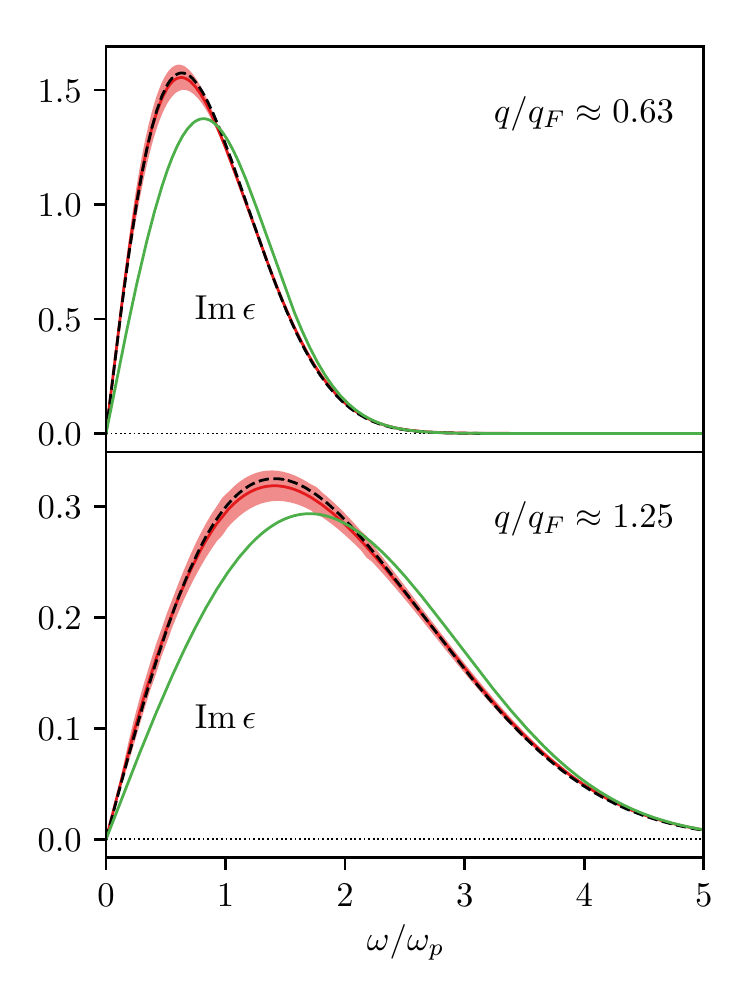}
    \caption{Top (bottom): Simulation results for the dielectric function $\epsilon(\vec{q},\omega)$ of the strongly correlated electron gas at $r_s=8$ ($r_s=2$) and $\theta=1$ for two wave numbers. \textbf{Left}: Real part of the dielectric function and imaginary part of the inverse dielectric function. \textbf{Right}: imaginary part of the inverse dielectric function. The peak of -Im$\epsilon^{-1}$ [and of $S(\vec{q},\omega)$] is in the vicinity of the second root of $\operatorname{Re}\epsilon$ (if roots exist, as in the upper figure). Green lines: RPA; red (dashed) lines: dynamic (static) PIMC results. 
    }
    \label{fig:eps-rs8}
\end{figure}

Using Eq.~(\ref{eq:eps-G}), the dynamic dielectric function is directly expressed by the local field correction to which we have access in our \textit{ab initio} simulations. Thus, it is straightforward to directly compare the RPA dielectric function to correlated  results that use either the static or dynamic LFC.

A first typical result for the dielectric function of the correlated electron gas is shown in Fig.~\ref{fig:eps-rs8}, for the cases of $r_s=2$ (bottom row) and $r_s=8$ (top row) and $\theta=1$. In the left (right) panel, we show the real (imaginary) part of the dielectric function for two wave numbers. At large frequencies, $\omega \gtrsim \omega_p$, the correlated results are in close agreement with the RPA. However, below $\omega_p$ deviations occur that increase with $r_s$. The peak of the imaginary part narrows and shifts to much lower frequencies. Due to the Kramers-Kronig relations, the same trend is observed for the real part. The statistical uncertainty of the reconstruction of $G(q,\omega)$ leads to an uncertainty in the region of the peak of Im$\,\epsilon$ that is indicated by the red band. Interestingly, the static approximation is very close to the full dynamic results, at the present parameters. In the left column of Fig.~\ref{fig:eps-rs8}, we also show the imaginary part of the inverse dielectric function, -Im$\,\epsilon^{-1}$, which is proportional to the dynamic structure factor, cf. Eq.~(\ref{eq:s-def}). For the case of $r_s=8$ at the lower wave number (top left figure), its peak is close to the larger zero of the real part of $\epsilon$, whereas for $r_s=2$ no zeroes of the real part of $\epsilon$ exist at these $q$-values.

\begin{figure}
    \centering
    \includegraphics[width=0.8\textwidth]{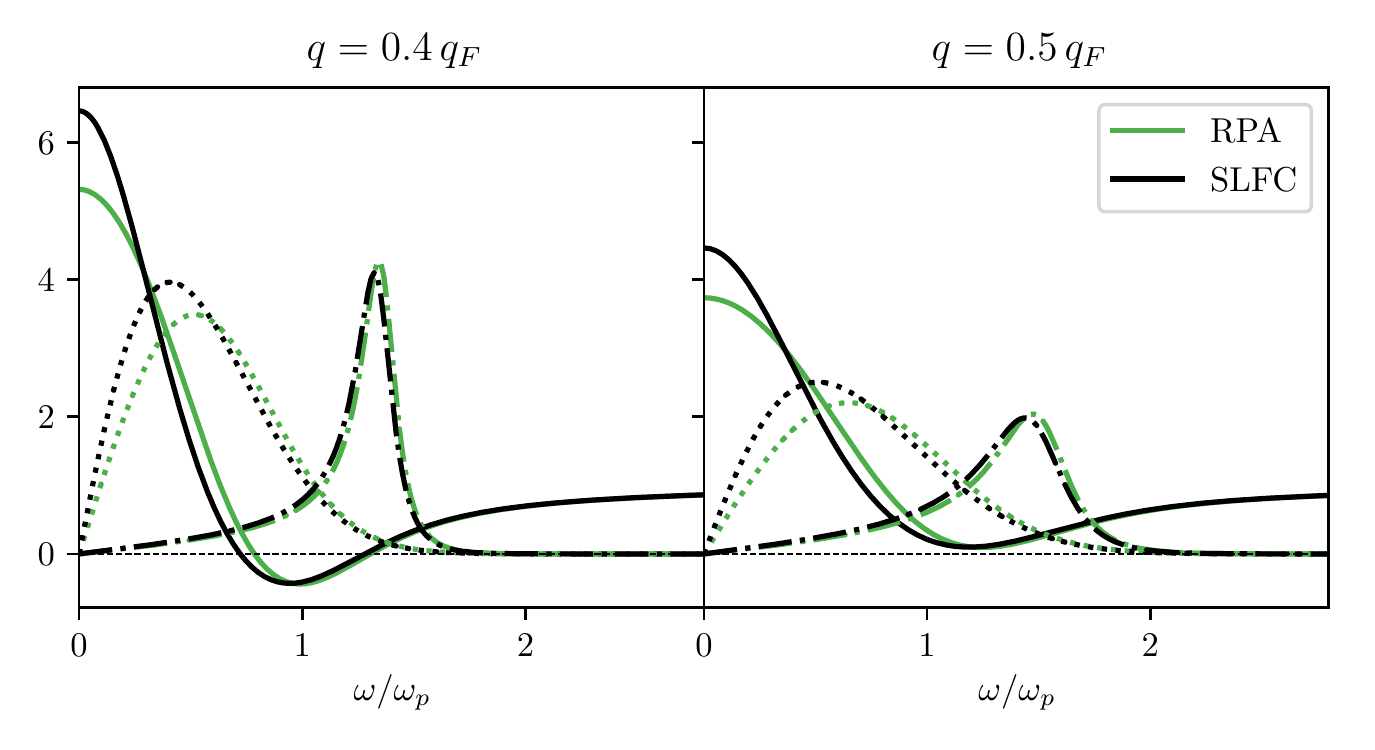}
    \caption{Real (solid line) and imaginary (dotted line) part of the dielectric function $\epsilon(q,\omega)$ for $\theta=1$ and $r_s=2$, for two wavenumbers. The results correspond to the values of the complex dielectric function for the same cases, shown in Fig.~\ref{fig:eps_komplex}, on the real frequency axis. Additionally, $-\operatorname{Im}\epsilon^{-1}$ is shown by the dash-dotted lines.}
    \label{fig:eps-rs2}
\end{figure}

The wavenumber dependence of the dielectric function for the case of $r_s=2$ is explored more in detail in Fig.~\ref{fig:eps-rs2}, for smaller $q$ than in Fig.~\ref{fig:eps-rs8}. Here we include only the result using the SLFC in addition to the RPA, because the difference to the full dynamic result is very small.
The existence of zeroes of Re $\epsilon$ on the real frequency axis sensitively depends on the wave number: for small wavenumbers, Re $\epsilon$ has two zeroes, but with increasing $q$, the zeroes vanish as in the examples of wave numbers $q=0.4 q_F$ and $0.5 q_F$ in Fig.~\ref{fig:eps-rs2}. During this transition, the value of the imaginary part of $\epsilon$ at the upper zero of the real part and thus the plasmon damping $\gamma(q)$ [Eq.~(\ref{eq:real-damping})] increase drastically. Hence, the peak of -Im $\epsilon^{-1}$ [and, with it, the peak of $S(q,\omega)$, cf. Eq.~(\ref{eq:s-def})] broadens strongly.
However, this transition is clearly beyond the validity of the weak damping approximation for the plasmon, ~(\ref{eq:real-dispersion}) and (\ref{eq:real-damping}), and requires to consider improved approximations that were discussed in Sec.~\ref{ss:dispersion} and \ref{ss:ac}. We will study these effects on the plasmon dispersion below. 
%


\subsection{Perturbation results for the RPA plasmon dispersion}\label{ss:results-dispersion}


%
As we discussed in Sec.~\ref{ss:dispersion} the approximation (\ref{eq:real-dispersion}) applies only for sufficiently weak damping. There we also discussed a straightforward way to relax this restriction. 
\begin{figure}
    \centering
    \includegraphics[width=\linewidth]{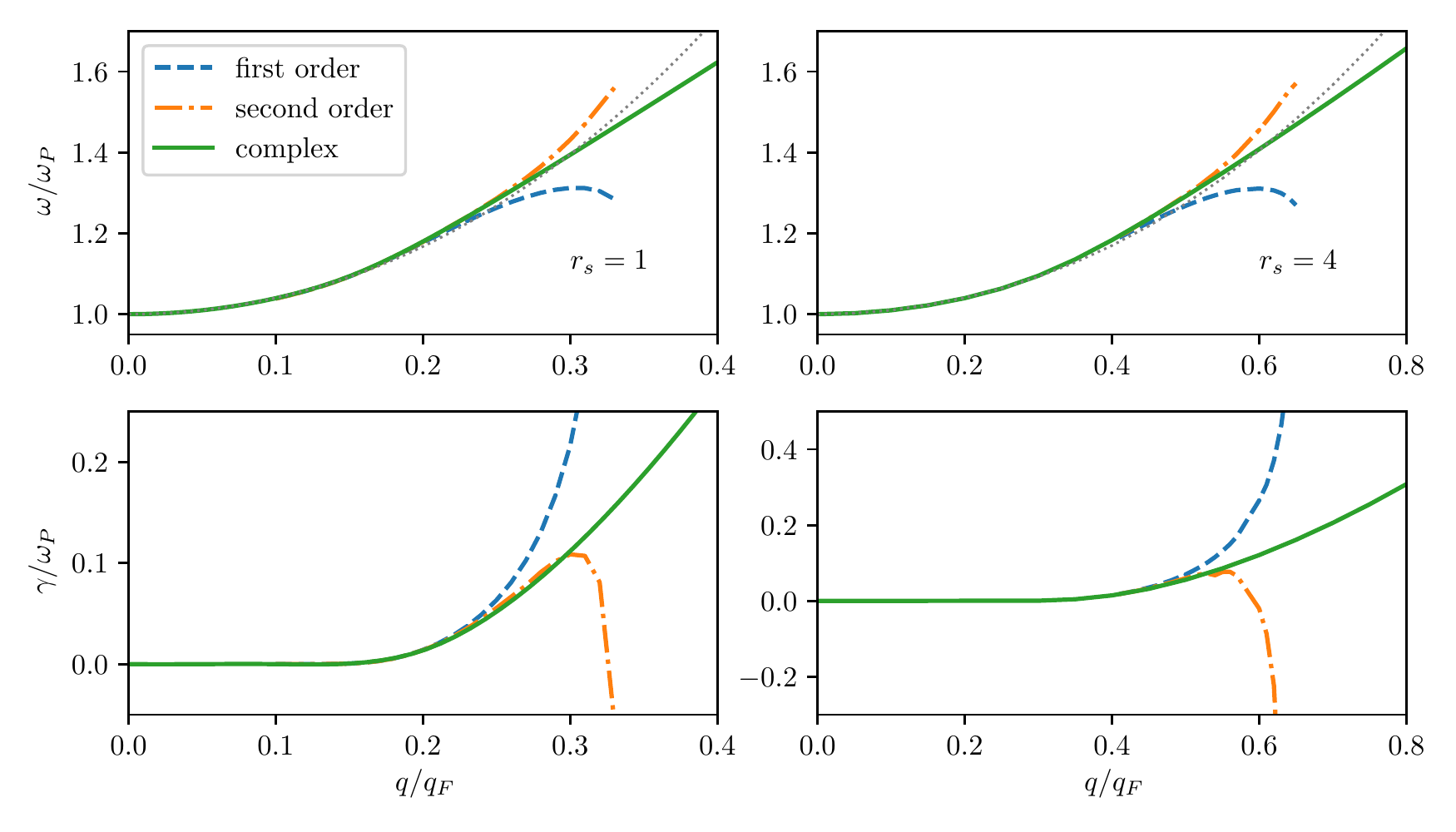}
    \caption{RPA-plasmon dispersion (top) and damping (bottom) for $\theta=1$ and $r_s=1$ (left) and $r_s=4$ (right): comparison of the complex dispersion relation (full green line)  to the small damping approximation, Eqs.~(\ref{eq:real-dispersion}) and (\ref{eq:real-damping}) -- blue dashed lines -- and next order expansion result (\ref{eq:disp-expansion-re-im-parts}) -- dash-dotted orange lines. Additionally, eq.~(\ref{eq:RPA_dispersion-finite-t-param}) is shown by the black dotted lines. Note that the solutions revealed by the Taylor expansion exist only up to a maximum wave number up to which the dispersion curves are drawn. The complex dispersion solution exists up to about $q/q_F\approx 1.0$, for $r_s=1$ and $q/q_F \approx 2.0$, for $r_s=4$.}
    \label{fig:taylor}
\end{figure}
To this end we have improved the weak damping  approximation, Eqs. (\ref{eq:real-dispersion}) and (\ref{eq:real-damping}) by extending the Taylor expansion of the dielectric function to third order in $|\gamma|/\omega$. The result is given by Eqs.~(\ref{eq:disp-expansion-re-im-parts}), for details see Appendix B. 
The results of the weak damping approximation and of the second order Taylor expansion [using the iterative procedure, cf. Appendix] are compared in Fig.~\ref{fig:taylor} for a moderate temperature, $\theta=1$, and two densities, $r_s=1$ and $r_s=4$. The figure confirms that the weak damping approximation,  Eqs. (\ref{eq:real-dispersion}) and (\ref{eq:real-damping}), agrees well with the next order of the expansion, Eqs.~(\ref{eq:disp-expansion-re-im-parts}), 
for small and moderate wavenumbers. Deviations start growing for $q/q_F \gtrsim 0.25$, for $r_s=1$, and $q/q_F \gtrsim 0.5$, for $r_s=4$. Notice that, at these wavenumbers, the ratio $|\gamma(q)|/\omega(q)$ is below $0.1$, but nevertheless the expansion already breaks down. Since it is not clear up to what $q$-values the higher order expansion, Eqs.~(\ref{eq:disp-expansion-re-im-parts}), will yield an improvement, we also include, in Fig.~\ref{fig:taylor} the full solution of the complex dispersion relation, Eq.~(\ref{eq:complex-dispersion}), because this is the true benchmark for the expansions. This requires to perform an analytic continuation of the dielectric function, as was demonstrated in Sec.~\ref{ss:ac}. 

The results of the analytic continuation, for the parameters of Fig.~\ref{fig:taylor}, are shown by the green lines [the numerical results will be discussed below, in Sec.~\ref{ss:ac-results}]. The comparison confirms that the higher order expansion is more accurate than the lowest order one and is valid up to larger wavenumbers than the latter, i.e. up to $q/q_F \approx 0.3$, for $r_s=1$, and up to $q/q_F \approx 0.6$, for $r_s=4$. The main conclusion from this comparison is that, even though the weak damping condition is well fulfilled [even at these wavenumbers, $|\gamma(q)|/\omega(q) \lesssim 0.2$], the approximation Re $\epsilon=0$ -- which is the commonly used condition for plasma oscillations -- fails badly. Not only are the values for the plasmon frequency wrong and, even more so, for the damping (except for small wavenumbers), the weak damping approximation also makes grossly incorrect predictions for the wavenumber range where plasmons exist.
We will return to this issue in Sec.~\ref{ss:dispersion-correlated}.

%


\subsection{Analytic continuation of the dielectric function}\label{ss:ac-results}
\begin{figure*}
    \centering
    \includegraphics[width=0.7\textwidth]{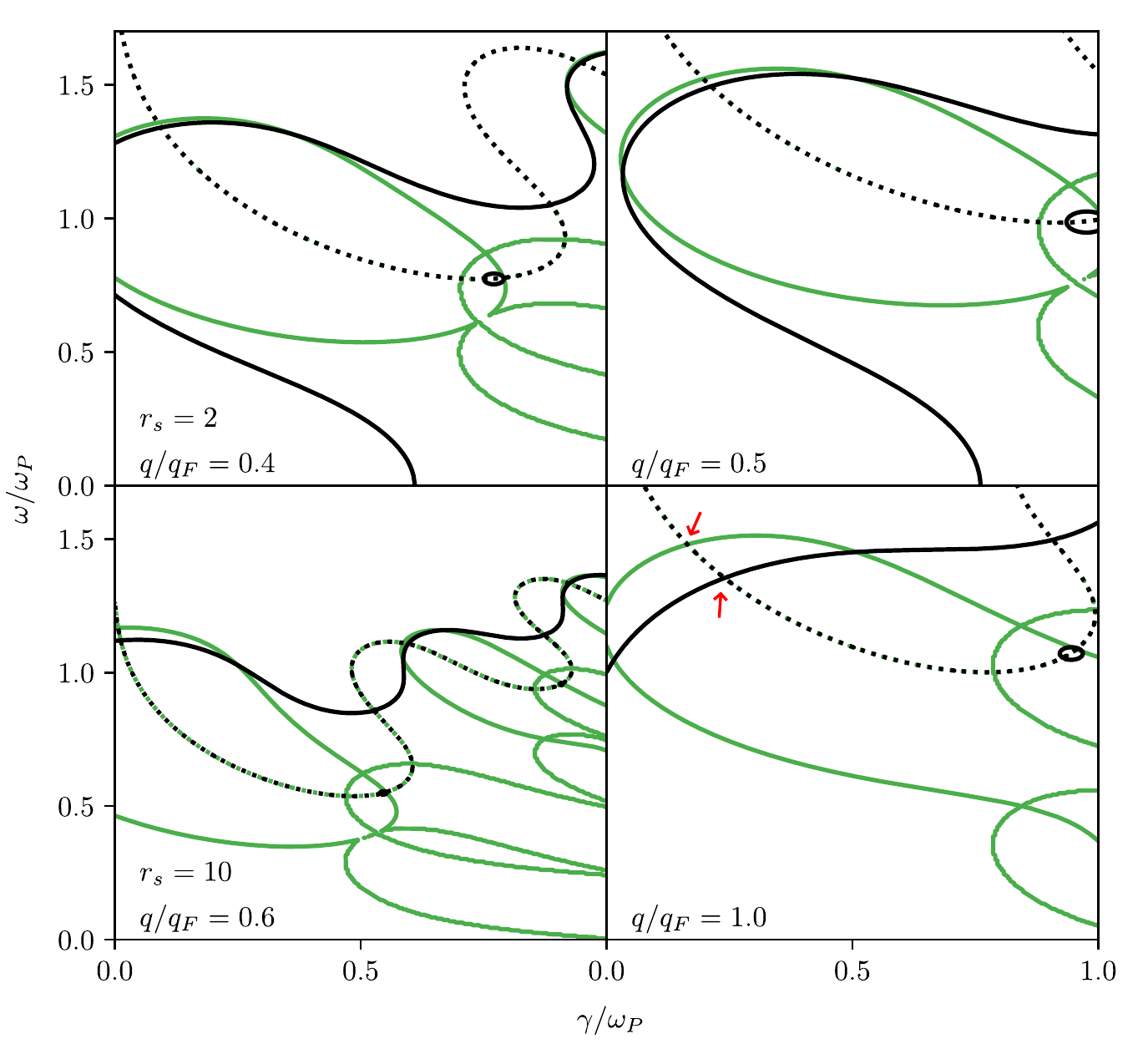}
    \caption{Analytic continuation of the dielectric function $\epsilon(q,\hat \omega)$ into the complex frequency plane, $\hat \omega=\omega -i\gamma$, for $\theta=1$ and two wavenumbers. {\bf Top}: $r_s=2$, {\bf Bottom}: $r_s=10$.
    The solid (dotted) lines represent zeroes of the real (imaginary) part, respectively. The black (green) lines are the SLFC (RPA) results. Note that, since $G(q,\omega=0)$ is purely real, using the SLFC has no influence on the position of the zeros of $\operatorname{Im}\epsilon(q,\omega)$, cf. Eq.~(\ref{eq:eps-G}). Plasmon frequency and damping can be identified from the intersections, where $\operatorname{Im}\epsilon(\hat \omega)=\operatorname{Re}\epsilon(\hat \omega)=0$ (see small red arrows in the bottom right panel. Only the intersection with the smallest damping has physical significance). Above a certain wave vector $q$, no roots of $\operatorname{Re}\epsilon$ exist on the real axis (top right panel), in agreement with Fig.~\ref{fig:eps-rs2}.}
    \label{fig:eps_komplex}
\end{figure*}
Due to the convergence problems of the weak damping expansion, it is important to improve the computation of the plasmon dispersion and damping by avoiding any weak damping ansatz and Taylor expansion. This is indeed possible, by resorting to the solution of the complex dispersion relation, Eq.~(\ref{eq:complex-dispersion}), i.e. by performing the analytic continuation of the dielectric function, as we discussed in Sec.~\ref{ss:dispersion}. 
We have carried out this procedure for both, the RPA and the static LFC (SLFC) approximation, Eq.~(\ref{eq:complex-eps-Gstatic}). The result of the analytic continuation, for the same parameters and two wave numbers of Fig.~\ref{fig:eps-rs2}, is plotted in the top row of Fig.~\ref{fig:eps_komplex}. This figure shows iso-lines of the zeroes of the real and imaginary parts of the dielectric function in the complex frequency space (full and dotted lines, respectively). Zeroes of the real part on the real axis are clearly visible for $q=0.4 q_F$, but no real zeroes survive for $q=0.5 q_F$ in agreement with Fig.~\ref{fig:eps-rs2}. At the same time, even for $q=0.5 q_F$, complex zeroes $\hat \omega(q)$, i.e. crossings of the full and dotted lines, exist. In the left panel ($q=0.4q_F$) the plasmon is located at 
$\hat \omega/\omega_p \approx 1.36 -i 0.09$ which is close the solution of the weak damping approximation. In the right part ($q=0.5q_F$) the plasmon parameters are $\hat \omega/\omega_p \approx 1.52 -i 0.2$.
%
There exist further crossings of the real and imaginary parts located at higher imaginary parts of the frequency. These excitations are strongly damped and therefore suppressed.

A second case of the analytic continuation, for the larger coupling of $r_s=10$ and $\theta=1$, is presented in the bottom row of Fig.~\ref{fig:eps_komplex}. As in the previous case, collective mode solutions in the complex plane extend to much larger wave numbers than the weak damping approximation predicts. Again the static LFC data are at lower frequency and higher damping than the RPA plasmon. This is particularly clearly visible at larger wavenumbers (right column). This figure also shows another peculiarity of the correlated dielectric function: for small wave numbers, the real part of $\epsilon$ has only a single branch Re $\epsilon(\hat \omega)=0$. This is not an artifact of the analytic continuation but a correlation effect that is visible also in the behavior of Re $ \epsilon$ as a function of real frequencies. A further analysis of this effect is given in Ref.~\cite{hamann_prb_20}. 
%


\subsection{\textit{Ab intio} plasmon dispersion of the correlated electron gas}
\label{ss:dispersion-correlated}
%
%
So far, we have considered the dispersion of plasma oscillations $\omega(q)$ and their damping $\gamma(q)$ focusing on the RPA. Having the correlated dielectric function available, cf. Sec.~\ref{ss:results-epsilon}, we can now extend this analysis to the correlated electron gas at finite temperature. Due to the close agreement of the static and dynamic approximations, we will only use $\epsilon^{\rm SLFC}$ to evaluate the correlated plasmon dispersion. 

\begin{figure}
    \centering
    \includegraphics[width=0.99\textwidth]{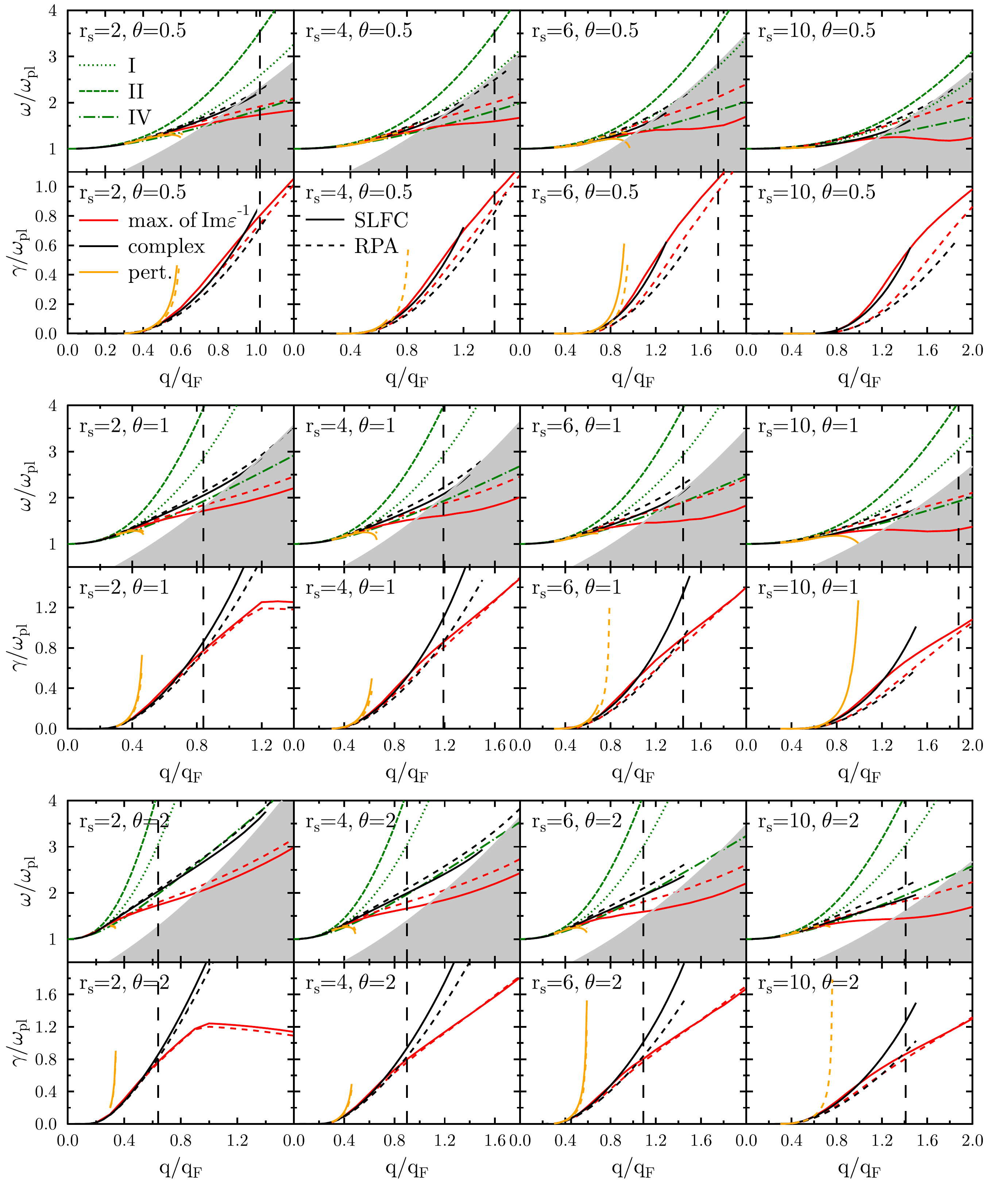}
    \caption{Plasmon dispersion and damping for a wide range of densities and temperatures. Black solid line (long dashed): analytical continuation using the SLFC (RPA); Red solid line (long dashed): Maximum of Im$\epsilon^{-1}\ofqw$ on the real axis; Orange solid line (long dashed): perturbation expansion for the dispersion of the SLFC (RPA) dielectric function according to Eq. (\ref{eq:real-damping}); green dots (dashed line, dash-dotted line): RPA plasmon dispersion for the weak damping case labelled as in Fig.~\ref{fig:RPA_dispersion}.  The vertical black dashed lines denote $\alpha=1$. The grey shaded area is the pair continuum.}
    \label{fig:dispersion_and_damping_all}
\end{figure}

Figure~\ref{fig:dispersion_and_damping_all}
summarizes the results for the plasmon dispersion and damping over a broad range of densities and temperatures corresponding to $2\le r_s \le 10$ and $0.5\le \theta \le 2$.  We include the results from the weak damping approximation and from the analytical continuation of both, $\epsilon^{\rm RPA}$ and $\epsilon^{\rm SLFC}$, and also the analytical approximations (\ref{eq:RPA_dispersion-finite-t}) and (\ref{eq:thiele-dispersion}) for the RPA dispersion. The main observations on the complex behavior are as follows.
\begin{enumerate}
    \item The general trend is that of steeper slopes of the dispersion with increased temperature and reduced slope with decreasing density. This overall trend is expected from the prefactor of the $q^2$ term in the RPA dispersion which is given by the thermal velocity, in the classical case, and the Fermi velocity, at strong degeneracy, for details see Sec.~\ref{ss:results-dispersion-wd}. This trend is reproduced well by all computational approaches.
    \item The inclusion of correlation effects via the SLFC shifts the plasmon to lower values, compared to the RPA, and reduces the slope (smaller prefactors of $q^2$ and $q^4$ terms). Further, as expected, the SLFC results are systematically more strongly damped than the RPA data because they include, in addition to Landau damping also correlation induced damping effects. 
    \item All approaches agree very well for small wave numbers where local field effects are negligible and the RPA collective modes exist and are accurate. Thus, for sake of efficiency and accuracy, the analytical low-$q$ series expansions I to IV (green lines) are to be preferred in this regime, with ``I'' corresponding to the parametrization (\ref{eq:RPA_dispersion-finite-t-param}), being the most accurate one.
    \item  The perturbative approaches that are based on the weak damping expansion and use Eqs. (\ref{eq:real-damping}) \& (\ref{eq:disp-expansion-re-im-parts}) allow for a direct inclusion of correlation effects via the SLFC and are, therefore, more useful in a broader regime, in particular, for larger $r_s$-values, cf. the yellow lines showing the RPA and SLFC results. For small wavenumbers, where the result for the damping is very small, these expansions turn out to be rather accurate, as compared to the analytical continuation. In contrast, the sudden increase of $|\gamma(q)|$ indicates the breakdown of these expansions.
    \item  However, the wave number range for which this is the case cannot be easily estimated beforehand, see the results in section \ref{ss:results-dispersion} and Fig.~\ref{fig:taylor}. In the presented range, it is found that one could use expansion IV (Eq.~(\ref{eq:RPA_dispersion-finite-t}), no $q^4$ terms) for an estimate of the plasmon location for higher temperatures.  Near the one-particle continuum (at low T), ``I'' [Eq.~(\ref{eq:RPA_dispersion-finite-t})] is to be preferred. 
    \item In accordance with the complex dispersion relation, Eq.~(\ref{eq:dispersion}), the method of analytical continuation (black lines) is the most accurate approach to the plasmon dispersion. It also yields solutions for intermediate wave numbers, where the perturbative approaches fail. At the same time for sufficiently large wavenumbers, the complex approach to the dispersion also ceases to find collective modes. Interestingly, the termination is observed close to the point where the dispersion $\omega(q)$ enters the pair continuum. 
    \item We also observe striking 
deviations of the location of the maximum of -Im$\epsilon^{-1}$ from the plasmon dispersion before the line $\alpha=1$, which gives a rough limit to the dominance of collective effects in the dielectric function, and also before the pair continuum is reached. We will discuss the significance of the fact below, in Sec.~\ref{ss:dispersion-vs-s}.

\end{enumerate}

We conclude that the perturbation approach (weak damping expansion, orange lines) and complex theory (black lines) give good agreement for the damping, for RPA and SLFC both, in the low-q range where the former is valid. On the other hand, the FWHM of the plasmon peak of -Im$\epsilon^{-1}$ on the real axis is in good agreement with the damping derived from the complex method. This can obviously only be the case for parameters for which a plasmon as such  exists and shows the limits of interpreting the peak of -Im$\epsilon^{-1}$ and its FWHM as  ``plasmon'' or ``plasmon damping''. Naturally, as soon as the peak width becomes of the order of the mode frequency, as is observed e.g. in the case $\theta=2, r_s=10$, in the bottom right part of Fig.~\ref{fig:dispersion_and_damping_all}, 
the interpretation as a collective excitation is not appropriate. We discusse this issue in more detail in the next section.


\subsection{Comparing the complex plasmon dispersion to the peak of the dynamic structure factor}\label{ss:dispersion-vs-s}
In Sec.~\ref{ss:sampling} we discussed how to obtain \textit{ab initio} PIMC data for the dynamic structure factor of the correlated electron gas. Even though this function is of prime importance for comparison with Thomson scattering data of warm dense matter, the physical interpretation of the results -- even for jellium -- is not trivial. The reason is that $S(q,\omega)$ is due to a number of different processes, including single-particle (particle-hole excitations) and collective plasma oscillations, e.g. \cite{thiele_pre_08,redmer_glenzer_2009}. 
From the relation to the dielectric function, Eq.~(\ref{eq:s-def}), it is clear that sharp peaks of $S(q,\omega)$ at small wave numbers most likely correspond to a collective mode. However, in the general case, a broad peak of $S$ is made from a mix of collective and single particle effect and thus does not directly yield information about the plasmon dispersion. Since correlation effects lead to an additional broadening of the peaks, this correspondence between plasmons and $S(q,\omega)$ becomes even more difficult. Having \textit{ab initio results} for both, $S(q,\omega)$ and $\epsilon(q,\omega)$, available, we are able to answer this question rigorously, for the first time. 

Before analyzing the numerical results, let us recall the commonly used criteria for a separation of collective and single-particle excitations. 
\begin{description}
\item[A.] At zero temperature, the boundary is given by the pair continuum that is shown in Fig.~\ref{fig:dispersion_and_damping_all} by the grey shaded area. Of course, at finite temperature, the pair continuum has merely qualitative relevance. 
\item[B.] The second criterion for collective excitations is the existence of zeroes of the dispersion relation. If no solutions exist at a given wave number, the excitations are expected to be entirely of single-particle nature. At the same time we have seen, that the existence condition of solutions strongly differs for the weak damping approximation and for the solution of the complex dispersion relation, where obviously the latter constitutes the most accurate result. 
\item[C.] Finally, a third qualitative separation of the two types of processes is often performed on the basis of the so-called scattering parameter~\cite{redmer_glenzer_2009}
\begin{equation}
\alpha=\frac{1}{q\lambda_s}\,,
\label{eq:alpha-parameter}
\end{equation}
where $\lambda_s$ is the screening length: for $\alpha \gtrsim 1$ ($\alpha < 1$) collective (single-particle) processes dominate the response of the plasma to the radiation of wavenumber $q$, as was already noted by Bohm and Gross \cite{bohm-gross_49}. The reason is that, for wavelengths exceeding the screening length all particles inside of a sphere of radius $\lambda_s$ will be excited simultaneously (``collectively''). Thus, Eq.~(\ref{eq:alpha-parameter}) predicts the existence of a critical wave number, $q_{\rm max}=1/\lambda_s$ beyond which no collective excitations exist. 
\end{description}
The existence of a maximum wavenumber for collective modes is in full qualitative agreement with the trends predicted by the other two criteria. 
It is, therefore, interesting to perform also a quantitative comparison of the criteria A--C over the relevant range of densities and temperatures. Such an analysis has already been performed in Fig.~\ref{fig:dispersion_and_damping_all}, and more detailed results will be given below in Fig.~\ref{fig:dispersion_skw_epsilon}. For the evaluation of the scattering parameter,  Eq.~(\ref{eq:alpha-parameter}), we use the stating long wavelength limit of the RPA polarization, as is described in Appendix C.

Let us now discuss the data displayed in Figs.~\ref{fig:dispersion_and_damping_all} and~\ref{fig:dispersion_skw_epsilon} with respect to the distinction between single particle and collective effects.  
First, the wave number $q_{\rm max}$ following from $\alpha=1$ is shown in all plots, cf. the vertical black dashed lines. Second, the pair continuum is shown by the grey shaded area. Complex zeroes of the dispersion relation have been plotted for all wavenumbers for which solutions of the corresponding dispersion relations exist. Remarkably, even for finite temperature, complex zeroes cease to exist in the vicinity of the grey areas (pair continuum). The estimation using $\alpha=1$ agrees excellently with the other two criteria for low temperatures and small $r_s$, whereas for higher temperatures, it predicts a significantly too small wavenumber $q_{\rm max}$. This is particularly striking at small $r_s$, e.g. the bottom left part of Fig.~\ref{fig:dispersion_and_damping_all}. Interestingly, for larger $r_s$, the criterion $\alpha=1$ works better again.
Of course, there is only a single strict criterion which is the existence of solutions of the complex dispersion relation (\ref{eq:complex-dispersion}) that follows from the analytical continuation of the dielectric function. 

Thus, Figure \ref{fig:dispersion_and_damping_all} allows one to verify the reliability of the other criteria and approximations. In particular, by comparing with the complex dispersion relation for $\epsilon^{\rm RPA}$, we conclude that use of the RPA static screening length in Eq.~(\ref{eq:alpha-parameter}) is most accurate for a particular $r_s$-value, at a given temperature: For $\Theta=0.5$, for $r_s\approx 2$, for $\Theta=1.0$, for $r_s\approx 6$, and for $\Theta=2.0$, for $r_s\approx 10$. For smaller (larger) values of $r_s$, at the same temperature, the screening length $r_s$ is overestimated (underestimated). Note that this behavior of the screening length is not due to the neglect of correlations but reflects a deficiency of the oversimplified criterion (\ref{eq:alpha-t0}).

After having discussed the wave number range where collective electronic plasma oscillations exist under WDM conditions, let us now turn to the dynamic structure factor and analyze how reliably it captures plasmons.
To this end, we compare in Figs.~\ref{fig:dispersion_and_damping_all} \& \ref{fig:dispersion_skw_epsilon} the results for the plasmon dispersion that follow from the complex zeroes of the dielectric function (black lines) to the data for the imaginary part of the inverse of the dielectric function (red lines, -Im$\epsilon^{-1})$ and to data for the dynamic structure factor $S\ofqw$ (blue lines) where, for the latter two cases, the position and FWHM of the peak are being used. This comparison is again performed for two sets of approximations -- the RPA, $\epsilon^{\rm RPA}$, and the static local field correction,  $\epsilon^{\rm SLFC}$, respectively.
The peak positions of $S(q,\omega)$ and -Im$\epsilon^{-1}\ofqw$ agree well with the plasmon dispersion obtained from the complex dielectric function for small wave numbers, both, for RPA and the correlated result following from the static LFC. However, when the wave number increases and reaches the wider vicinity of any of the criterias A--C, the peak of -Im$\epsilon^{-1}(q,\omega)$ and even more so the peak of the dynamic structure factor quickly fall below the plasmon dispersion. The effect is particularly strong for the correlated (SLFC) result when $r_s$ and thus the coupling increases. In particular, for $r_s\gtrsim 6$ the peak positions of -Im$\epsilon^{-1}(q,\omega)$ and $S\ofqw$ even decrease with $q$ in a finite wavenumber range (for $r_s=10$ this range is approximately between $k_F$ and $2 k_F$). This was first observed in Ref.~\cite{dornheim_prl_18} where it was interpreted as indication of a negative plasmon dispersion. As the width of the peak of $S(q,\omega)$ in that range of wavenumbers is of the same order as the frequency $\omega(q)$ of the peak, one  would rather expect an overdamped oscillation. 
Even though this effect should be observable in XRTS experiments and is an exciting correlation effect in WDM, it is, however, mainly due to single-particle excitations. This conclusion can now be clearly made, based on our analysis of the complex plasmon dispersion that reveals a strictly monotonic increase of $\omega(q)$ where the solution ceases to exist at $q\approx 1.4$ ($q\approx 1.5$) for SLFC (for RPA), in the case of $r_s=4$. For $r_s=10$, the critical wavenumbers are correspondingly $q\approx 1.43$ for both, SLFC and RPA.

In general, the value of the FWHM of $S\ofqw$ and -Im$\epsilon^{-1}\ofqw$ is only equal to the damping of the plasmons for very small wavenumbers. As for the location of the plasmon that was discussed above, deviations between plasmon damping and FWHM start to appear early in such a way that the FWHM can be larger or smaller than the damping of the plasmon. If the FWHM is larger than the plasmon damping from the complex theory, we take that as an indicator for (correlated) single particle effects to contribute to the signal. When, for large wavenumbers, the FWHM is lower than the plasmon damping, we find an overdamped state with mostly single particle excitations contributing to the spectrum.
\begin{figure}
    \centering
    \includegraphics[width=0.75\textwidth]{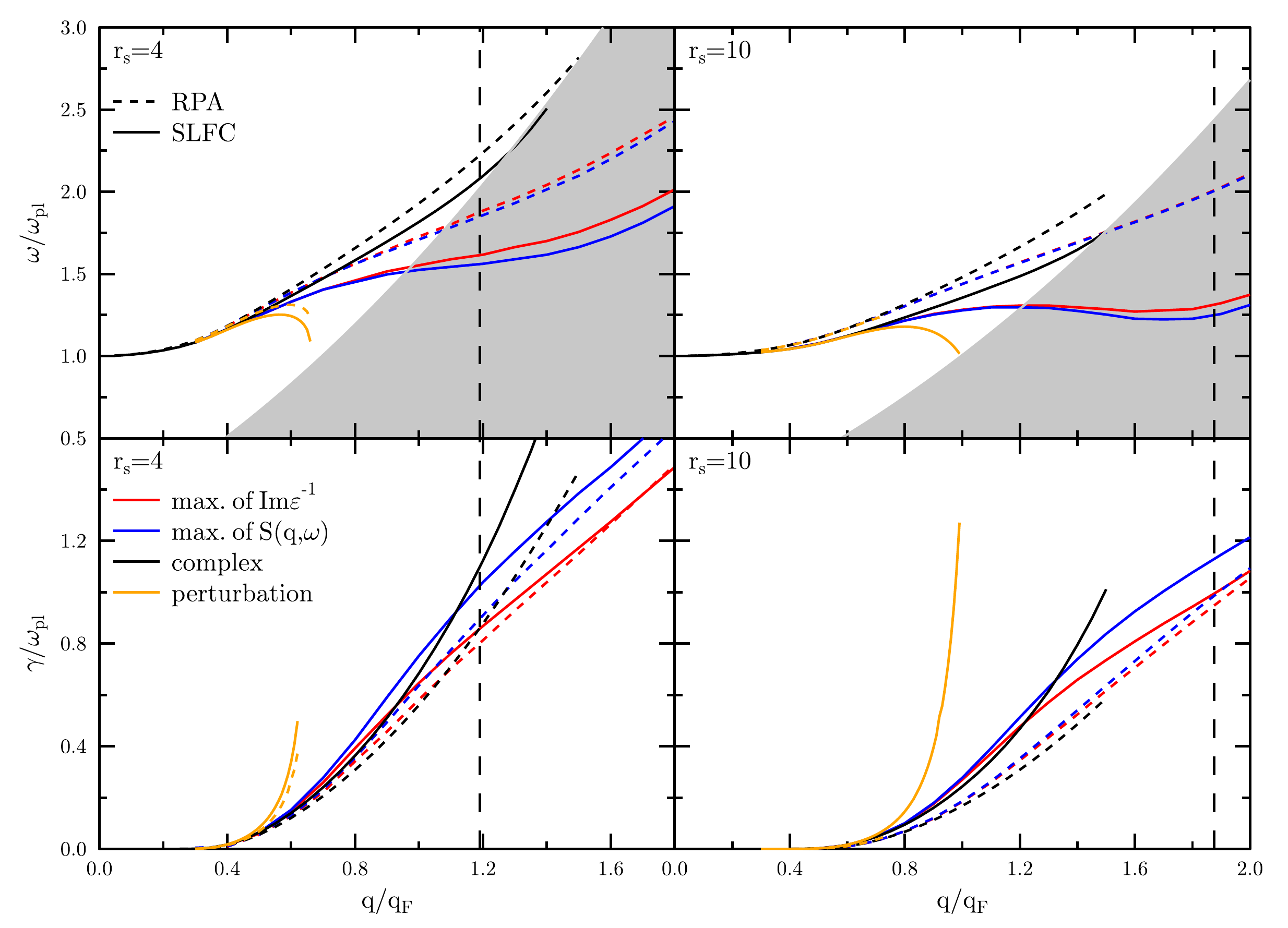}
    \caption{Plasmon dispersion (top) and damping (bottom) for $\theta=1$ and two densities, obtained from the analytic continuation of the dielectric function (black) in RPA (long dashed) and using the static LFC (solid line).
    Orange lines: perturbation expansion on the real axis. Red lines: peak position and width (FWHM) of the corresponding Im$\epsilon^{-1}(q,\omega)$. 
    blue lines: peak position and width (FWHM) of the corresponding dynamic structure factors.
   }
    \label{fig:dispersion_skw_epsilon}
\end{figure}

Thus, the procedure to extract plasmon dispersions from an XRTS experiment should never be based on the dynamic structure factor alone, except for small wavenumbers and narrow peaks. In general, one should use, in addition the spectral function which is the imaginary part of the inverse dielectric function as well as the dielectric function itself and solve the dispersion relation. Since Im$\epsilon^{-1}$ naturally contains all possible excitations, not just plasmons, to extract the collective excitations one has to analyze the dielectric function.


\section{Summary and Discussion}\label{s:summary}
In this paper we have extended the \textit{ab initio} quantum Monte Carlo approach to the dynamic structure factor, that was presented in Ref.~\cite{dornheim_prl_18}, to the dynamic dielectric function. In particular, the \textit{ab initio} results for the dynamic and static local field correction have allowed us to obtain unbiased accurate results for the dispersion of longitudinal electronic plasma oscillations that fully include correlation and finite temperature effects. To explore the effect of exchange and correlations, we performed a detailed comparison of SLFC results to the random phase approximation (RPA).

To have reference results, we revisited the commonly used RPA dispersion relation, $\epsilon^{\rm RPA}(q,\omega)=0$, and compared different analytical parametrizations of $\omega(q)$. The most accurate one was found to be the ground state result of Ferrell, Eq.~(\ref{eq:ferrell-dispersion}), which we extended to arbitrary finite temperatures,  and for which we presented an accurate analytical parametrization given by Eq.~(\ref{eq:RPA_dispersion-finite-t-param}).

In addition to the analytical parametrizations for the RPA plasmon dispersion, we explored several numerical methods of obtaining the plasmon dispersion and damping: perturbation theory (with respect to the damping) for the zeroes of the dielectric function on the real frequency axis, higher order perturbation theory results, and 
the analytical continuation of the retarded dielectric function into the lower frequency half plane. Finally we also performed a direct determination of the location and width of the plasmon from the peaks of the plasmon spectral function, -Im$\eps^{-1}$.

The true plasmon dispersion and damping are determined by the complex dispersion relation, requiring one to solve simultaneously the two equations, Re$\epsilon=$Im$\epsilon=0$, with the complex solution $\hat \omega(q)=\omega(q)-i\gamma(q)$. The spectral function reflects all aspects of the longitudinal excitations, not just plasmons, and the position, FWHM and shape of its peak can be directly compared to the XRTS measurements.

However, the interpretation of the experimental signal in terms of collective or single-particle excitations, requires an independent evaluation of the plasmon dispersion. 
%
Here all of the presented methods give identical results for very small wavenumbers, deep in the collective regime. In this case, the perturbation theory has clear advantages as it does not suffer much from numerical noise as is the case for the complex method or the spectral functions. For mixed regimes of collective and single-particle excitations, we observed that a separation based on the scattering parameter $\alpha$ can be very inaccurate. Instead,  
the analytic continuation method is clearly the best in extracting the plasmon excitations and its wave number range. 

The present first \textit{ab initio} results for the  plasmon dispersion, including correlation effects, will be particularly important for WDM experiments at elevated temperatures, $T \sim E_F$, i.e. in the plasma phase. The biggest correlation effects should occur for comparatively low densities where $r_s \sim 4 \dots 10$, and wavenumbers of the order of $1k_F\dots 2k_F$. A suitable candidate could be hydrogen.

\section*{Acknowledgements}
This work is supported by the German Science Foundation (DFG) via grant BO1366-15.
TD acknowledges support by the Center of Advanced Systems Understanding (CASUS) which is financed by Germany’s Federal Ministry of Education and Research (BMBF) and by the Saxon Ministry for Science, Culture and Tourism (SMWK) with tax funds on the basis of the budget approved by the Saxon State Parliament.
Zh. Moldabekov is thankful for the funding from the German Academic Exchange Service (DAAD). This work
has been supported by the Grant  AP08052503 of Ministry of Education and Science of the Republic of Kazakhstan. 
All PIMC calculations were carried out on the clusters \emph{hypnos} and \emph{hemera} at Helmholtz-Zentrum Dresden-Rossendorf (HZDR), the computing centre of Kiel university, at the Norddeutscher Verbund f\"ur Hoch- und H\"ochstleistungsrechnen (HLRN) under grant shp00015, and on a Bull Cluster at the Center for Information Services and High Performance Computing (ZIH) at Technische Universit\"at Dresden.

\appendix
\renewcommand{\thefigure}{A\arabic{figure}}
\setcounter{figure}{0}

\section*{Appendix A: Analytical parametrization of the RPA plasmon dispersion relation}\label{s:appendix}
In this Appendix we present details on the analytical plasmon dispersion for a uniform electron gas at finite temperature that were discussed in Sec.~\ref{ss:results-dispersion}. We consider the regime with $\omega\gg \hbar q^2/(2m)$ and $\omega\gg qv_F$. 
In this limit, the expansion of the real part of the dielectric function in RPA has the following form \cite{arista-brandt_84}:
\begin{equation}
\mbox{Re}\,\epsilon(q,\omega) = 1 -\frac{\omega_p^2}{\omega^2} \left( 1+ \frac{\langle v^2\rangle}{\omega^2}\,q^2
+ \frac{\langle v^4\rangle}{\omega^4}\,q^4 +\frac{\hbar^2}{4m^2}\frac{q^4}{\omega^2}+\orderof\left( \tilde{q}^{\,6}\right) \right),
\label{eq:rpa_expan1}
\end{equation}
where $\tilde{q}=(q/q_F)(\omega_p/\omega)$.
%
In Eq.~(\ref{eq:rpa_expan1}),  the  velocity moments can be expressed in terms of the Fermi integrals of order $\nu$, $I_\nu$, as
\begin{equation}
    \langle v^{\alpha}\rangle=\frac{3}{2} v_F^{\alpha}\, \theta^{(\alpha+3)/2}\, I_{(\alpha+1)/2}(\eta),
\end{equation}
where $\eta=\mu/k_B T$ and $v_F$ is the Fermi velocity.

From the condition of the plasmon resonance, in the weak damping approximation, Re$\,\epsilon(q,\omega) = 0$, and  Eq.~(\ref{eq:rpa_expan1}), the dispersion relation follows: 
\begin{align}
\frac{\omega^2(q)}{\omega_p^2} &= 1  + \frac{\langle v^2\rangle}{\omega^2(q)}\, q^2 + \left\{ \frac{\langle v^4\rangle}{\omega^2(q)} + \frac{\hbar^2}{4m^2}\right\}\frac{q^4}{\omega^2(q)}
 +\orderof\left( \tilde{q}^{\,6}\right).
\label{eq:RPA_dispersion_main}
\end{align}
%
which contains the plasmon dispersion also on the r.h.s., so we proceed by iteration. 
The first iteration is obtained by substituting $\omega(q)=\omega_p$ into the right hand side of Eq. (\ref{eq:RPA_dispersion_main})~\cite{arista-brandt_84, thiele_pre_08}:
\begin{align}
\frac{\omega^2(q)}{\omega_p^2} &= 1  + \frac{\langle v^2\rangle}{\omega_p^2}\, q^2 + \left\{ \frac{\langle v^4\rangle}{\omega_p^2} + \frac{\hbar^2}{4m^2}\right\}\frac{q^4}{\omega_p^2}
+\orderof\left( \tilde{q}^{\,6}\right).
\label{eq:RPA_dispersion0}
\end{align}
The second iteration follows from  substituting the result for $\omega(q)$ from Eq. (\ref{eq:RPA_dispersion0}) into the right hand side of Eq. (\ref{eq:RPA_dispersion_main}):
\begin{equation}
\frac{\omega^2(q)}{\omega_p^2} = 1  + \frac{\langle v^2\rangle}{\omega_p^2 B(q,\omega)}\, q^2  \\ 
+\left\{ \frac{\langle v^4\rangle}{\omega_p^2 B(q,\omega)} + \frac{\hbar^2}{4m^2}\right\}\frac{q^4}{\omega_p^2 B(q,\omega)}+\orderof\left( \tilde{q}^{\,6}\right),
\label{eq:RPA_dispersion1}
\end{equation}

\begin{equation}
    B(q)= 1+ \frac{\langle v^2\rangle}{\omega_p^2}\, q^2+\frac{\langle v^4\rangle}{\omega_p^4}\,q^4 ++\frac{\hbar^2}{4m^2}\frac{q^4}{\omega_p^2}+\orderof\left( \tilde{q}^{\,6}\right).
\end{equation}
On the same level of approximation as Eq.~(\ref{eq:rpa_expan1}), we expand $B^{-1}(q)$ as:
\begin{equation}
    \frac{1}{B(q)}= 1-\frac{\langle v^2\rangle}{\omega_p^2}\, q^2+\left\{1- \frac{\langle v^4\rangle}{\omega_p^2} - \frac{\hbar^2}{4m^2}\right\}\frac{q^4}{\omega_p^2}+\orderof\left( \tilde{q}^{\,6}\right).
    \label{eq:B1}
\end{equation}
Substituting Eq.~(\ref{eq:B1}) into  Eq.~(\ref{eq:RPA_dispersion1}) we arrive at \cite{ferrell_57, dandrea_86}:
 \begin{align}
        \frac{\omega^2(q)}{\omega_p^2} &= 1  + \frac{\langle v^2\rangle}{\omega_p^2}\, q^2 + \left\{ \frac{\left(\Delta v^2\right)^2}{\omega^2_p} + \frac{\hbar^2}{4m^2}\right\}\frac{q^4}{\omega_p^2}
        +\orderof\left( \tilde{q}^{\,6}\right),
        \label{eq:RPA_dispersion2}
    \end{align}
where  $\Delta v^2 = [\langle v^4 \rangle - \langle v^2 \rangle^2]^{1/2}$. 
On the level of approximation of Eq.~(\ref{eq:rpa_expan1}), further iterations do not change the result for the dispersion relation,  Eq.~(\ref{eq:RPA_dispersion2}), which constitutes the most accurate long-wavelength expansion of the RPA dispersion for the finite temperature electron gas. 

In Fig.~\ref{fig:RPA_dispersion} we compare various analytical approximations for the plasmon dispersion that were presented here and in Sec. \ref{ss:results-dispersion}, for the case $\theta=1$ and $r_s=4$. This figure clearly shows that the benchmark result Eq.~(\ref{eq:RPA_dispersion2}) [line 3] is very distinct from the simpler  analytical approximations. Similar deviations are observed for other temperatures and densities.

It is, therefore, important for an accurate description of the plasmon dispersion of the uniform electron gas, to use the full result, Eq. (\ref{eq:RPA_dispersion2}).
Evaluating the second and fourth moments of the velocity with the finite temperature Fermi distribution allows us to find the parametrization (\ref{eq:RPA_dispersion-finite-t-param}) of the dispersion, that was presented in the main text where, for the coefficients in front of the $q^2$ and $q^4$ terms, $B_2(r_s,\theta)$ and $B_4(r_s,\theta)$, we have obtained the following analytical approximations,
\begin{equation*}
    B_2(r_s,\theta)=\frac{6.783}{r_s} \frac{\left(\frac{4}{25}+\theta^2\right)^{1/2}}{1-0.14\left[\exp(-\theta)-\exp(-3.68 ~\theta)\right]},
\end{equation*}
and 
\begin{equation*}
    B_4(r_s,\theta)=\frac{8.7633}{r_s^2}\left(1+6.4875~\theta^{\frac{2}{1.16}}\right)^{1.16}
    -\left[ B_2(r_s,\theta)\right]^2
    +\frac{1.1305}{r_s}.
\end{equation*}
The parametrization (\ref{eq:RPA_dispersion-finite-t-param}) with the coefficients $B_2(r_s,\theta)$ and $B_4(r_s,\theta)$ agrees with the exact numerical results with a precision better than $3\%$ in the entire range of $\theta$, $r_s$, and $q$ and also reproduces the exact analytical limits at $\theta\ll 1$ and $\theta\gg 1$.\\ 

\section*{Appendix B: Plasmon dispersion relation for moderate damping}\label{s:appendix_b}
Here we extend the weak damping result for the plasmon dispersion and damping, Eqs.~(\ref{eq:real-dispersion}) and (\ref{eq:real-damping}), to stronger damping. To this end, we extend the Taylor expansion of the complex dispersion relation to terms of order $(\gamma/\omega)^3$:
\begin{align}
    0&= \text{Re}\,\epsilon(\omega) -i\gamma \text{Re}\,\epsilon'(\omega) + \frac{(-i\gamma)^2}{2!}\text{Re}\,\epsilon''(\omega)
    + \frac{(-i\gamma)^3}{3!}\text{Re}\,\epsilon'''(\omega) 
    \nonumber\\
&   \quad +i \text{Im}\,\epsilon(\omega) 
 +\gamma \text{Im}\,\epsilon'(\omega) 
 -\frac{i\gamma^2}{2}\text{Im}\,\epsilon''(\omega)+ \dots\,,
 \label{ap:disp-expansion}
\end{align}
There are several ways to solve this equation. 
\subsection*{Iterative solution}
The first is an iterative solution, for a fixed $q$.
\begin{description}
\item[iteration 1:] The first approximation for the frequency, $\omega_1(q)$, follows from the lowest order to the real part of Eq.~(\ref{ap:disp-expansion}):\\  $\text{Re}\,\epsilon(\omega_1)=0 $, and coincides with approximation (\ref{eq:real-dispersion}).
\item[iteration 2:] Inserting this result into the imaginary part of Eq.~(\ref{ap:disp-expansion}) yields [we denote $d/d\omega$ by a prime] $\gamma_1=\gamma(\omega_1)$:
\begin{align}
\gamma_1 = \frac{\text{Im}\,\epsilon(\omega_1)}{\text{Re}\,\epsilon'(\omega_1)} \,,
\nonumber
\end{align}
which is the previous result (\ref{eq:real-damping}) that is improved in the following.
\item[iteration 3:] Inserting $\gamma_1$ into the real part of Eq.~(\ref{ap:disp-expansion}) yields $\omega_2(q)$:
\begin{align}
0 = \text{Re}\,\epsilon(\omega_2) -\frac{\gamma_1^2}{2}\text{Re}\,\epsilon''(\omega_2)+\gamma_1 \text{Im}\,\epsilon'(\omega_2)\,.
\label{ap:dispersion2}
\end{align}
\item[iteration 4:] Inserting $\omega_2$ into the imaginary part of Eq.~(\ref{ap:disp-expansion}) yields $\gamma_2(q)$:
\begin{align}
\gamma_2 = \frac{\text{Im}\,\epsilon(\omega_2)}{\text{Re}\,\epsilon'(\omega_2)} + \frac{\gamma_1^3}{3!}\text{Re}\,\epsilon'''(\omega_2)
- \frac{\gamma_1^2}{2!}\text{Im}\,\epsilon''(\omega_2)\,,
\label{ap:damping2}
\end{align}
\end{description}
and so on. While this iterative approach significantly improves the result for the dispersion and damping, it has the disadvantage that the existence condition for solutions $\omega(q)$ is determined by the first iteration, i.e. by the weak damping dispersion relation.

\subsection*{Selfconsistent solution}
The iterative procedure can be avoided by solving Eq.~(\ref{ap:disp-expansion}) directly, simultaneously for $\omega(q)$ and $\gamma(q)$. This leads to the system of two coupled equations for the real and imaginary parts of Eq.~(\ref{ap:disp-expansion}) that yields two real functions $\omega(q)$ and $\gamma(q)$
\begin{align}
    0&= \text{Re}\,\epsilon(\omega) - \frac{\gamma^2}{2!}\text{Re}\,\epsilon''(\omega)
     +\gamma \text{Im}\,\epsilon'(\omega) \,,
\nonumber     
 \\
    0&= \gamma \text{Re}\,\epsilon'(\omega) 
 - \text{Im}\,\epsilon(\omega) 
- \frac{\gamma^3}{3!}\text{Re}\,\epsilon'''(\omega)  +\frac{\gamma^2}{2}\text{Im}\,\epsilon''(\omega)\,.
 \label{ap:disp-expansion-re-im-parts}
\end{align}
If the damping is moderate, one can restrict the second equation to the first two terms on the right because the last two terms are of third order in the damping.


\section*{Appendix C: Scattering parameter and screening parameter}\label{s:appendix_b}
\begin{figure}[h]
    \centering
    \includegraphics[width=0.45\textwidth]{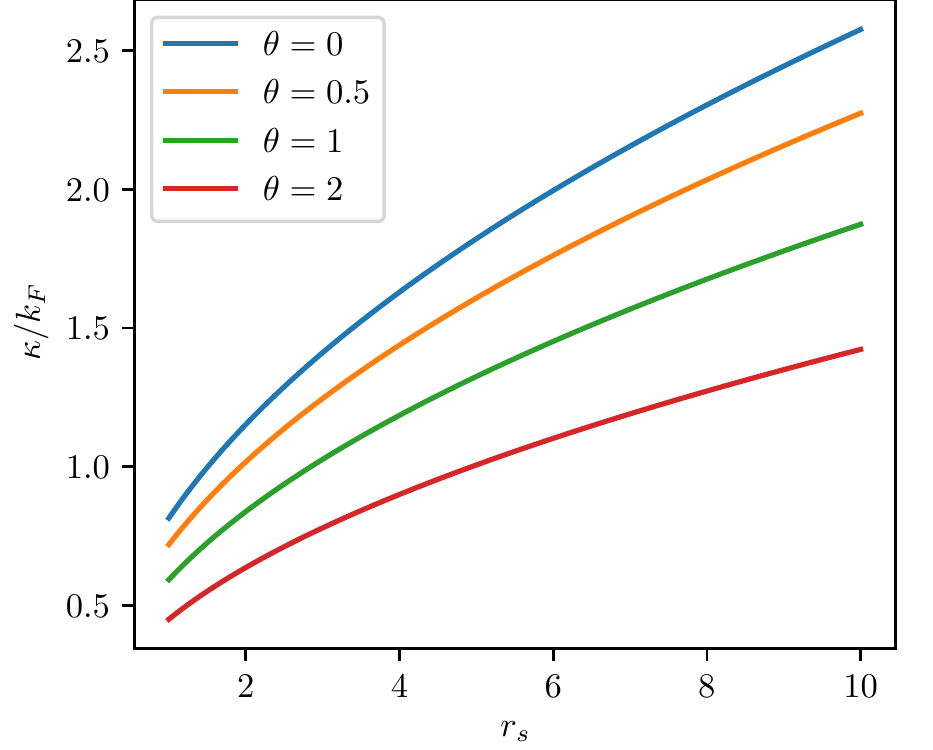}
    \caption{Inverse screening parameter $\kappa=\lambda_s^{-1}$ -- the long wavelength limit of the static RPA polarization -- as a function of the density parameter, at different temperatures. The $\theta=0$ result corresponds to the usual Thomas-Fermi screening, Eq.~(\ref{eq:screening-tf-t0}). The model of the scattering parameter $\alpha$, Eq.~(\ref{eq:alpha-parameter}), predicts that collective modes exist only up to a wavenumber $q_{\rm max}=\kappa$.}
    \label{fig:screening}
\end{figure}
The simplest approximation for the screening length is that for an ideal Fermi gas which, in the ground state, is given by the Thomas-Fermi length
\begin{align}
    \lambda^2_{TF}(T=0) &= \frac{4}{3}\frac{1}{0.88}\frac{1}{r_s}\frac{1}{q_F^2}\
    = \frac{1}{3}\frac{v_F^2}{\omega^2_p}\,,
    \label{eq:screening-tf-t0}\\
    \frac{1}{\alpha(T=0)} &= 
    \frac{2}{\sqrt{3}}\frac{1}{\sqrt{0.88 r_s}}\frac{q}{q_F}\,.
\label{eq:alpha-t0}
\end{align}
For finite temperature this result is generalized by computing the static long wavelength limit of the RPA polarization  function $\Pi$, Eq.~(\ref{eq:response}). The result is shown in Fig.~\ref{fig:screening}. For zero temperature, we recover the analytical result (\ref{eq:screening-tf-t0}) whereas, with increasing temperature, the screening length (the screening parameter $\kappa=\lambda_s^{-1}$) increases (decreases). For very high temperatures, the result is identical to the Debye screening. Thus, the maximum wavenumber for collective modes, $q_{\rm max}=\kappa$ predicted by criterion C can be directly read off from Fig.~\ref{fig:screening}.


\bibliography{ref,mb-ref}

\end{document}